\documentclass[12pt]{article}
\pdfoutput=1 

\usepackage{jheppub} 

\usepackage[T1]{fontenc} 
\usepackage{amsmath,physics}
\usepackage{comment}
\usepackage{bbm}
\usepackage{mathtools}
\usepackage{tikz-cd}
\usepackage[toc,page]{appendix}
\usepackage{etoolbox}
\usepackage{tabularx}

\BeforeBeginEnvironment{appendices}{\clearpage}
\AfterEndEnvironment{appendices}{\clearpage}
\AfterEndEnvironment{appendices}{\pretocmd{\section}{\clearpage}{}{}}{}
\AfterEndEnvironment{appendices}{\pretocmd{\subsubsection}{\clearpage}{}{}}{}

\usepackage{graphicx,float,caption,subcaption,xcolor}
\usepackage{hyperref}
\hypersetup{
    colorlinks=true,
    linkcolor=blue,
    filecolor=magenta,      
    urlcolor=cyan,
    pdftitle={Penrose limits and Double Copies},
    pdfpagemode=FullScreen,
    }

\graphicspath{{./img/}} 
\usepackage{multirow}

\usepackage{bbm}

\title{Expanding on the double copy \\ in null Fermi coordinates}
\author[a]{Samarth Chawla,}
\author[b]{Kwinten Fransen,}
\author[a]{Cynthia Keeler}

\affiliation[a]{Department of Physics, Arizona State University, Tempe, AZ 85281, USA}
\affiliation[b]{Walter Burke Institute for Theoretical Physics, California Institute of Technology, Pasadena, CA 91125, USA}

\emailAdd{samarthc@asu.edu}
\emailAdd{kfransen@caltech.edu}
\emailAdd{keelerc@asu.edu}

\abstract{We propose a Weyl classical double copy for a Fermi normal coordinate expansion around null geodesics. To leading order in this ``Penrose expansion'', we recover a previously proposed double copy of the Penrose limit. For spacetimes with an exact double copy, this Penrose limit double copy is extended to all orders. For spacetimes without such a double copy, generic obstructions appear at second subleading order. We thus argue that for \emph{any} spacetime, near \emph{any} null geodesic there is a classical double copy structure at least up to first subleading order in the Penrose expansion. Finally, we point out a difficulty in identifying an appropriate flat space to extend our results to the Kerr-Schild double copy, related to the generic incompatibility between Kerr-Schild and Penrose-G\"uven gauge.}

\begin{document} 
\maketitle
\flushbottom
 
\newpage

\section{Introduction}

The double copy has been a driving force in the study and calculation of amplitudes since the observation by
\cite{Bern:2008qj} of a ``color-kinematics duality''; expressing how relations between gauge theory amplitudes due to the color structure can be mirrored in their kinematical pieces. In addition to allowing the calculation of gravitational amplitudes using tools
developed for calculating Yang-Mills amplitudes
\cite{Bern:2019nnu,Bern:2019crd,Bern:2020uwk,Bern:2021yeh,Bern:2022jvn}, the discovery of the double copy also led
to a deeper examination of shared structure between various gauge and gravity theories that was previously unknown.
See \cite{Bern:2019prr,Kosower:2022yvp,White:2024pve} for recent reviews.

A deceptively simple question is whether or not the double copy structure of amplitudes can also be seen directly in the classical limit of field theories. Early attempts to relate the
amplitudes double copy to classical solutions are summarized in chapter 8 of \cite{Bern:2019prr}. The
first \textit{exact} formulation of the double copy for classical solutions was described in \cite{Monteiro:2014cda} for Kerr-Schild spacetimes. The Weyl double copy proposal for
certain algebraically special spacetimes
\cite{Luna:2018dpt} was another important milestone in the study of the exact classical double copy. This Weyl double copy proposal has the advantage of being formulated in terms of gauge-invariant objects. 

A classical double copy structure has now been established for a large number of examples
\cite{Luna:2015paa,Luna:2017dtq,Ridgway:2015fdl,Adamo:2017nia,Bahjat-Abbas:2017htu,CarrilloGonzalez:2018ejf,Ilderton:2018lsf,Gurses:2018ckx,Carrillo-Gonzalez:2017iyj,Lee:2018gxc,Lescano:2020nve,Lescano:2021ooe,Bah:2019sda,Goldberger:2019xef,Kim:2019jwm,Luna:2020adi,Easson:2020esh,Berman:2018hwd,Alkac:2021seh,Mkrtchyan:2022ulc,Adamo:2022rob,Alfonsi:2020lub,Adamo:2020qru,Gonzo:2021drq,Luna:2018dpt,Keeler:2020rcv,Easson:2021asd,Easson:2022zoh,Godazgar:2020zbv,White:2020sfn,Chacon:2021wbr,Chacon:2021lox,Han:2022mze,Alawadhi:2020jrv,Godazgar:2021iae,Armstrong-Williams:2023ssz,Armstrong-Williams:2024bog}. However, it remains unclear how general these double copy descriptions of spacetime are and, often, to which, if any, amplitude double copy they are related. Moreover, while the amplitudes double copy began as a perturbative relation between scattering amplitudes on flat space there have since been further generalizations to all-order results and perturbative amplitudes on curved backgrounds \cite{Adamo:2017nia,Adamo:2023fbj,Sivaramakrishnan:2021srm,
	Herderschee:2022ntr,Cheung:2022pdk,Ilderton:2024oly,Cheung:2022mix}. Correspondingly, the classical double copy has a growing
body of examples where the various gauge theory fields are put on a curved spacetime instead of on an auxiliary flat space
\cite{Bahjat-Abbas:2017htu,Carrillo-Gonzalez:2017iyj,Gurses:2018ckx,Prabhu:2020avf,Alkac:2021bav,Han:2022ubu,Han:2022mze,Didenko:2022qxq,He:2023iew,Kent:2024mow,Zhao:2024wtn}.

A pressing question in relation to the momentum-space double copy for amplitudes is the mechanism by which they can give rise to a local position-space double copy of classical solutions. More generally, a double copy that
is local in momentum space should lead to a convolutional double copy in position space
\cite{Anastasiou:2014qba,LopesCardoso:2018xes,Anastasiou:2018rdx,Luna:2020adi,Godazgar:2022gfw}. As clarified in \cite{Luna:2022dxo} building on \cite{Monteiro:2021ztt}, a local,
position-space double copy is possible as a consequence of the algebraic speciality of the known examples. This reliance on algebraic speciality raises doubts on the formulation of equally successful double copy relations for more general spacetimes. 

Nevertheless there have been efforts to extend the scope of the classical double copy beyond special cases. One direction of research uses the Weyl double copy paradigm but finds a double copy structure in a
term-by-term expansion for metrics that are of more general algebraic type \cite{Keeler:2020rcv,Keeler:2024bdt}. Other proposals embrace more directly the non-local nature of the double copy by a detour via twistor space, but still find a Weyl double copy at the linearized level
\cite{White:2020sfn,Chacon:2021wbr,Chacon:2021lox} for general algebraic type. Alternatively, it is interesting to study the interface between algebraically special spacetimes and more general spacetimes by first
reducing the metric to special submanifolds, where a double copy relation is known to hold, and subsequently examining the double copy
structure as we move away from the submanifold.

This last approach was taken in \cite{Godazgar:2021iae} for asymptotically flat metrics in Bondi coordinates. In an expansion around null infinity, the
leading order pieces of the metric are of algebraic type N, to which the type N Weyl double
copy proposal of \cite{Godazgar:2020zbv} applies. Instead, in certain non-radiative cases, the leading asymptotic spacetime is of type D and the Weyl double copy of \cite{Luna:2018dpt} applies. In the examples studied by \cite{Godazgar:2021iae}, this asymptotic double copy structure breaks down at next-to-leading order except for special cases.

 We propose a similar study of the double copy, expanding the metric around
null geodesics instead \cite{penrose1976any,Blau:2006ar}. As already found in \cite{Chawla:2024mse}, in the exact limit of the metric at the null
geodesic a double copy description is always possible. In this paper we go beyond the exact
limit to look at the subleading terms in the expansion around a null geodesic, and explore how far into the
expansion the double copy structure persists. We keep much of our discussion as general as possible, which means
that our results may be applied near null geodesics in the interior of arbitrary spacetimes.

The plan of the rest of the paper is as follows. We first lay out the necessary background concerning the formalism
of the Weyl double copy in Section \ref{sec:background}. Section \ref{sec:overview} offers a high-level overview of our argument concerning the application of the Weyl double copy in an expansion around a null geodesic. Sections \ref{eqn:curvedlocal} and \ref{sec:normalneighborhoods} subsequently fill in all the technical details of the argument. Section
\ref{sec:kerrschild} examines whether the Kerr-Schild double copy can also be formulated in such an expansion,
primarily relying on the example of the Schwarzschild metric to illustrate the inherent complications of making a
Kerr-Schild version of the double copy work.

\paragraph{Conventions} We follow the conventions of Penrose and Rindler \cite{Penrose:1985bww}, except for the Riemann curvature which differs by a sign from those references. See also our
previous paper \cite{Chawla:2024mse}, which uses the same conventions as used here, and Appendix \ref{app:NP}.

\section{Weyl double copy in the Penrose expansion} 

\subsection{Background}\label{sec:background}
The Weyl double copy relation between the Weyl spinor $\Psi_{ABCD}$, the Maxwell spinor $f_{AB}$, and the ``zeroth copy'' scalar field $S$ is given by \cite{Luna:2018dpt}
\begin{equation} \label{WeylDCRelation} 
\Psi_{ABCD} = \frac{1}{S} f_{(AB} f_{CD)} \, .
\end{equation}
In terms of the Weyl and Maxwell scalars in an arbitrary frame, \eqref{WeylDCRelation} can be written as \cite{Godazgar:2021iae}
\begin{equation}\label{eqn:scalarsdcrelations}
\begin{aligned}
\Psi_0 = \, \frac{(f_0)^2}{S}\,, \quad
\Psi_1 = \, \frac{f_0\,f_1}{S} \,, \quad
\Psi_2 = \, \frac{f_0\, f_2 + 2 (f_1)^2}{S} \,, \quad
\Psi_3 =  \,\frac{f_1\, f_2}{S} \,, \quad
\Psi_4 = \, \frac{(f_2)^2}{S} \,.
\end{aligned} 
\end{equation}
In addition to the algebraic relations \eqref{WeylDCRelation}, or \eqref{eqn:scalarsdcrelations}, the heart of the
double copy proposal is that both the gravitational double copy as well as the electromagnetic and scalar single
and zeroth copy satisfy appropriate field equations. Exactly which set of field equations should be satisfied is
open to debate and various proposals have been made.

In this section, we will consider $S$ and $f_{AB}$ to be solutions to a modified massless scalar and electromagnetic field equations on the \emph{curved} background associated to $\Psi_{ABCD}$\footnote{On the other hand, even though the single and zeroth copy live on the curved background, they do not act as sources for the field equations satisfied by the geometry associated to $\Psi_{ABCD}$.}. In the next section, we will discuss and leverage an additional Kerr-Schild structure to analyze the possibility that $S$ and $f_{AB}$ are solutions to the scalar and electromagnetic field equations on an appropriately defined \emph{flat} spacetime\footnote{A double copy where the single copy satisfies curved field equations is sometimes referred as a ``type B'' double copy. In this terminology, when the gauge theory is defined on a flat background it is referred to as a ``type A'' double copy \cite{Bahjat-Abbas:2017htu,White:2024pve}.}. \\

Explicitly, the field equations we will impose for the double copy and the single copy are
\begin{equation}\label{eqn:ZRMeqns}
\nabla^{AA'}\Psi_{ABCD} = 0 \, , \quad \nabla^{AA'} f_{AB} = 0
\end{equation}
where, as emphasized above, the covariant derivatives refer to the curved spacetime in both cases. These are the
``zero-rest-mass equations'' \cite{Penrose:1985bww} which (together with their conjugates) represent the usual vacuum field equations for a spin two and spin one massless field respectively. \\

The scalar field equation is more subtle. The reason is that it is \emph{not} true in general that the zeroth copy satisfies the curved massless scalar field equation. An alternative is proposed in \cite{Godazgar:2021iae} in terms of the same massless scalar field equation with a source depending on $\Psi_2$ in a principal null frame. Instead, we propose 
\begin{equation}\label{eqn:eomscalarcurved}
\left(\square  +  \sqrt{2 \Psi_{ABCD} \Psi^{ABCD}/3}\right) S = 0\, .
\end{equation}
The scalar equation of motion \eqref{eqn:eomscalarcurved} is equivalent to the proposal of \cite{Godazgar:2021iae} for a Weyl double copy spacetime but has otherwise several advantages: (i) it is a homogeneous equation in $S$, (ii) it is independent of a preferred frame choice, and (iii) it readily reduces to the Fackerell-Ipser equation for a type D Weyl double copy spacetime \cite{Fackerell:1972hg}. We provide some details in appendix \ref{app:FIequation} but the upshot is that, for a type D spacetime, \eqref{eqn:eomscalarcurved} will be automatically satisfied if \eqref{WeylDCRelation} and \eqref{eqn:ZRMeqns} are satisfied. For a type N spacetime, \eqref{eqn:eomscalarcurved} just reduces to a massless scalar field equation.

\subsection{Overview and strategy}\label{sec:overview}
Before going into the details let us outline our strategy. We ultimately want to consider the relations \eqref{eqn:scalarsdcrelations}, together with the relevant field equations \eqref{eqn:ZRMeqns} and \eqref{eqn:eomscalarcurved}, expanded in Fermi null coordinates around a given null geodesic $\gamma$. By \cite{Chawla:2024mse}, we know that the leading order Penrose limit will have a double copy in arbitrary spacetimes that can be made consistent with an exact Weyl double copy if it is present. 

On the other hand, we are not aware of general expressions to obtain the structure of the metric directly in Fermi null coordinates to arbitrary order. While it seems algorithmically possible to provide such extensions this is likely impractical \cite{li1979expansions}. Therefore, we are instead led to start from adapted coordinates which do have a clear all order transformation to Fermi null coordinates, extending the leading order relation between Rosen and Brinkmann coordinates \cite{Blau:2006ar}. 

Starting from adapted coordinates, which exactly represent a patch of the full spacetime, we will readily find that if \eqref{eqn:scalarsdcrelations}, \eqref{eqn:ZRMeqns}, and \eqref{eqn:eomscalarcurved} are satisfied exactly, they will also be satisfied order-by-order. More interestingly, for an arbitrary spacetime which is not necessarily of the Weyl double copy form,  we can still try to solve  the relations \eqref{eqn:scalarsdcrelations}, \eqref{eqn:ZRMeqns}, and \eqref{eqn:eomscalarcurved} order-by-order from the bottom-up and we find that these equations can nevertheless be generically satisfied up to first order.

Finally, adapted coordinates are not globally well-defined in a normal neighborhood around the geodesic $\gamma$, the natural spacetime patch described by Fermi null coordinates. Therefore, after analyzing in detail the double copy structure in a stretch of $\gamma$ free of conjugate points, where adapted coordinates are defined, we study the transition between such patches in order to show that our results on the double copy of the Penrose expansion can be established ``globally'' in a null normal neighborhood of $\gamma$.

\subsection{Geodesic stretches free of conjugate points}\label{eqn:curvedlocal}

A stretch of the geodesic $\gamma$ free of conjugate points can be embedded into a null geodesic congruence for which we can construct the associated adapted coordinates
\begin{equation}\label{eqn:adapted}
ds^2 = 2 dU dV + D(x^{\mu}) dV^2 + 2 B_i(x^{\mu}) dV dX^i - C_{ij}(x^{\mu})dX^i dX^j \, ,
\end{equation}
with $D(x^{\mu})$, $B_i(x^{\mu})$, and $C_{ij}(x^{\mu})$ functions of all the coordinates $x^{\mu} = \left\lbrace U, V, X^i \right\rbrace$ and with indices $i$, $j$ running over two transverse directions ($i,j, \ldots \in \left\lbrace 1, 2 \right \rbrace$). By construction, the stretch of the null geodesic $\gamma$ captured by this coordinate system is located at $X^i = V = 0$. 

After choosing a transverse frame
\begin{equation}\label{eqn:adaptedframetransverse}
E_{(a)} = E^{i}_{(a)} \partial_i \, , \quad E^{i}_{(a)} C_{ij} E^{j}_{(b)} = \delta_{ab} \, , \quad C^{ik}C_{kj} = \delta^i_{j} \, ,
\end{equation}
an example of a full pseudo-orthonormal frame for the metric \eqref{eqn:adapted} is given by $\left\lbrace k, n,  E_{(a)} \right\rbrace$ (for $(a)\in \left\lbrace 1, 2 \right \rbrace$) where
\begin{equation}\label{eqn:adaptedframe}
\begin{aligned}
k^{\mu}\partial_{\mu} = \partial_U \, , \quad n^{\mu}\partial_{\mu} = \partial_V + B^{(a)}  E_{(a)}- \frac{1}{2}(B_{(a)}  B^{(a)} +D)\partial_U \, .
\end{aligned}
\end{equation}
with
\begin{equation}
B_{(a)} = B_i E^i_{(a)} \, , \quad B^{(a)} = B_i E^i_{(b)}\delta^{ab} \, .
\end{equation}

The adapted coordinates, or the underlying null geodesic congruence, single out the vector field $k$ by construction, otherwise our choice for \eqref{eqn:adaptedframe} was arbitrary. That is, the frame \eqref{eqn:adaptedframe} is only fixed up to spacetime-dependent Lorentz transformations preserving $k$. These transformations take the form
\begin{equation}\label{eqn:littlegroup}
E_{(a)} \to O^b{}_{a} E_{(b)} + c_{(a)} k \, , \quad n \to n + c_{(a)} \delta^{ab} E_{(b)} + c_{(a)}\delta^{ab} c_{(b)} k \, ,
\end{equation}
where $c_{(a)}$ are scalar fields and $O^b{}_{a}$ is a field of orthogonal matrices. Together, these transformation generate the massless ``little group'' in four dimensions $O(2) \ltimes \mathbb{R}^2$. The little group action can be used to transform to a new frame $\left\lbrace k, \hat{n},  \hat{E}_{(a)} \right\rbrace$, which is parallel transported along the stretch of the null geodesic $\gamma$ captured by the coordinate system.

The Penrose expansion corresponds to an expansion in $\epsilon \ll 1$ after setting $(U,V,X^i) \sim (\epsilon^0, \epsilon^2, \epsilon)$. On the assumption that the functions $D$, $B_i$, and $C_{ij}$ are finite and smooth in this limit, it implies a leading order scaling of the metric as $ds^2 \sim \epsilon^2$ and of the above frame as $\left\lbrace k^{\mu}, n^{\mu},  E^{\mu}_{(a)} \right\rbrace \sim \left\lbrace \epsilon^0, \epsilon^{-2},  \epsilon^{-1} \right\rbrace $. Said differently, up to a uniform scaling, the frame scales as: $\left\lbrace k^{\mu}, n^{\mu},  E^{\mu}_{(a)} \right\rbrace \to \epsilon^{-1} \left\lbrace \epsilon k^{\mu}, \epsilon^{-1} n^{\mu},  E^{\mu}_{(a)} \right\rbrace $. The relative scalings will thus align with the weights in the Geroch-Held-Penrose (GHP) formalism \cite{geroch1973space}, which determine the behavior of Newman-Penrose quantities exactly under such transformations of the frame. For instance, the GHP-weights of the Weyl scalars and Maxwell scalars of a test field on the background imply the ``peeling''-behaviors 
\begin{equation}\label{eqn:peeling}
\frac{\Psi_I}{\Psi_0} \sim \epsilon^I \, , \quad \frac{f_I}{f_0} \sim \epsilon^I \, ,
\end{equation}
with $I \in \left\lbrace 0,1, \ldots, 2s \right \rbrace$ with $s = 2$ for $\Psi_I$ and $s=1$ for $f_I$. See \cite{Kunze:2004qd} as well as Appendix \ref{app:Penroseexpansion} for more details on how the spin coefficients, Weyl scalars, and Maxwell scalars of a frame like \eqref{eqn:adaptedframe} scale\footnote{Note that \cite{Kunze:2004qd} uses slightly different notation and conventions.}. 

In order to have a well-defined tetrad in the plane wave limit $\epsilon \to 0$, the metric as well as the tetrad
must be appropriately rescaled. It would be proper to distinguish the rescaled metric and tetrad by
using different notation, such as
\begin{align} \label{} 
    \mathring{ds}^{2} &= \epsilon^{-2} ds^{2} \\
    \left( \mathring{k}^{\mu}, \mathring{n}^{\mu}, \mathring{E}^{\mu}_{(a)}\right) &= \left(k^{\mu},
    \epsilon^2 n^{\mu}, \epsilon E^{\mu}_{(a)}\right),
\end{align}
and so on. We relegate such careful distinction and further explanation to appendix \ref{app:Penroseexpansion}, and for the rest of this section we will simply drop the ring diacritic with the understanding
that all quantities have been appropriately rescaled.

To leading order in $\epsilon$, only $\Psi_0$ and $f_0$ contribute and these Weyl and Maxwell scalars characterize a plane gravitational wave and electromagnetic wave respectively corresponding to the Penrose-G\"uven limit of the full fields \cite{Gueven:2000ru}. Consider now the structure of the higher orders. Denote
\begin{equation}\label{eqn:pert}
\Psi_I = \sum^{\infty}_{k=0} \epsilon^k \Psi^{(k)}_I \, , \quad f_I = \sum^{\infty}_{k=0} \epsilon^k f^{(k)}_I \, , \quad S = \sum^{\infty}_{k=0} \epsilon^k S^{(k)} \, ,
\end{equation}
such that a superscript ${}^{(0)}$ indicates the leading Penrose limit order, which is of the form \cite{Chawla:2024mse}
\begin{equation}\label{eqn:eqnpenroselimit}
\Psi^{(0)}_{0} = 	\Psi^{(0)}_{0}(U) \, ,  \quad 	f^{(0)}_{0} = 	f^{(0)}_{0}(U) \, ,  \quad  S^{(0)} = S^{(0)}(U) \, , \quad	\Psi^{(0)}_{I>0} =  f^{(0)}_{I>0}  = 0 \, .
\end{equation}
Then, assuming the fields result from a smooth Penrose expansion of an exact set of fields, the  following polynomial form is expected at each order
\begin{align}
S^{(k)} &= \sum^{2n_v + \sum_i n_i =  k}_{n_i,n_v=0} a^{(k)}_{S,n_v, n_i}(U)  V^{n_v} \prod^2_{i=1} (X^i)^{n_i} \, , \label{eqn:Sexpansion} \\
f^{(k)}_I &= \sum^{2n_v + \sum_i n_i =  k-I}_{n_i,n_v=0} a^{(k)}_{f_I,n_v, n_i}(U)  V^{n_v} \prod^2_{i=1} (X^i)^{n_i} \, , \label{eqn:fexpansion} \\
\Psi^{(k)}_I &= \sum^{2n_v + \sum_i n_i =  k-I}_{n_i,n_v=0} a^{(k)}_{\Psi_I,n_v, n_i}(U)  V^{n_v} \prod^2_{i=1} (X^i)^{n_i} \, . \label{eqn:Psiexpansion}
\end{align}
Using these expansions, we can now perturbatively solve \eqref{eqn:scalarsdcrelations} together with \eqref{eqn:ZRMeqns} and \eqref{eqn:eomscalarcurved} for $f_I$ and $S$. Unless stated otherwise, we assume the non-degenerate case $S^{(0)} , \, 	f^{(0)}_{0}  , \, \Psi^{(0)}_{0} \neq 0$.

First, we find it useful to invert, at each order, the first three equations of \eqref{eqn:scalarsdcrelations} in order to express the Maxwell scalars $f^{(i)}_I$ in terms of the geometric data and the scalars  $S^{(i)}_I$, as well as lower order quantities\footnote{$f_0^{(k)}$ does not actually contribute to $f^{(k)}_1$ because $f^{(0)}_1 = 0$ and similarly, $f_0^{(k)}$ and $f_1^{(k)}$ do not contribute to $f_2^{(k)}$.}
\begin{equation}\label{eqn:fexpansiongeneral}
\begin{aligned}
f^{(k)}_0 &= \frac{1}{2f_0^{(0)}}\left(-\sum^{k-1}_{q=1}f^{(q)}_0f^{(k-q)}_0  +\sum^k_{q=0} S^{(q)}\Psi_0^{(k-q)}\right) \, ,\\
f^{(k)}_1 &= \frac{1}{f_0^{(0)}}\left(-\sum^{k}_{q=1}f^{(q)}_0f^{(k-q)}_1 +\sum^k_{q=0} S^{(q)}\Psi_1^{(k-q)}\right) \, ,  \\ 
f^{(k)}_2 &= \frac{1}{f_0^{(0)}}\left(-\sum^{k}_{q=1}f^{(q)}_0f^{(k-q)}_2-2\sum^k_{q=0}f^{(q)}_1f^{(k-q)}_1+\sum^k_{q=0} S^{(q)}\Psi_2^{(k-q)}\right) \, . \\ 
\end{aligned}
\end{equation}
By the leading order scalings of $\Psi_3$ and $\Psi_4$ (and consistently $f_1$ and $f_2$), the last two equations of \eqref{eqn:scalarsdcrelations} only appear as additional consistency conditions on \eqref{eqn:fexpansiongeneral} starting respectively from the third and fourth subleading order. We will discuss them in more detail when needed. 

Consider instead the equations of motion. After inserting \eqref{eqn:pert} and collecting order by order, we find the standard result that at each order the homogeneous equation for the relevant order is simply the zeroth order equation, but there is a source term depending on all lower orders. For the scalar equation of motion
\begin{equation}\label{eqn:Seqnspert}
\begin{aligned}
\square^{(0)}S^{(k)} = J_S^{(k)}[S^{(0)}, \ldots, S^{(k-1)}] \, ,
\end{aligned}
\end{equation}
while for the Maxwell scalars in the Newman-Penrose notation (see Appendix \ref{app:NP})
\begin{align}
D^{(0)} f^{(k)}_{1} - \overline{\delta}^{(0)} f^{(k)}_{0} + 2 \rho^{(0)} f^{(k)}_{1} &=  J^{(k)}_{f_1}[f^{0}_I, \ldots f^{k-1}_I], \label{eqn:feqnspert1} \\
D^{(0)} f^{(k)}_{2} - \overline{\delta}^{(0)} f^{(k)}_{1} +  \left(\rho^{(0)} - 2 \varepsilon^{(0)}\right) f^{(k)}_{2} &= J^{(i)}_{f_2}[f^{0}_I, \ldots f^{k-1}_I], \label{eqn:feqnspert2} \\
\delta^{(0)} f^{(k)}_{1} - \Delta^{(0)} f^{(k)}_{0}- \sigma^{(0)} f^{(k)}_{2}&=  J^{(k)}_{f_3}[f^{0}_I, \ldots f^{i-1}_I], \label{eqn:feqnspert3} \\
\delta^{(0)} f^{(k)}_{2} - \Delta^{(0)} f^{(k)}_{1} &=   J^{(k)}_{f_4}[f^{0}_I, \ldots f^{k-1}_I] . \label{eqn:feqnspert4}
\end{align}
The implications of the structure of the equations \eqref{eqn:Seqnspert} and \eqref{eqn:feqnspert1}-\eqref{eqn:feqnspert4} are two-fold: (i) at each order we have in principle the freedom to add a homogeneous solution and (ii) at each order it is straightforward to find a solution. 

Point (i) could complicate the analysis by the amount of freedom it provides in constructing solutions but this freedom is removed by the expected structure  \eqref{eqn:Sexpansion}-\eqref{eqn:Psiexpansion} of the solutions. Instead, point (ii) is just the statement that the zeroth order equations are the massless scalar and electromagnetic test field equations on a plane wave spacetime, which on account of its large isometry group has separable solutions to the wave equation.  

Using the metric in adapted coordinates \eqref{eqn:adapted}, the order-by-order massless scalar field equation \eqref{eqn:Seqnspert} is given more explicitly by 
\begin{equation}\label{eqn:Seqnspertexpl}
\begin{aligned}
\left( \frac{1}{\sqrt{C_{(\gamma)}}}\partial_U \sqrt{C_{(\gamma)}} \partial_V + \partial_V\partial_U  - C_{(\gamma)}^{ij}  \partial_i  \partial_j \right)S^{(k)} = J_S^{(k)} \, , \quad C_{(\gamma)} = C|_{X^i=V=0} \, .
\end{aligned}
\end{equation}
Here, it is important that $C_{(\gamma)}$ and $C_{(\gamma)}^{ij}$ depend only on $U$. For a general discussion of the formal solutions see Appendix \ref{app:leadingequations}. On the other hand, restricting to the ansatz \eqref{eqn:Sexpansion}, observe first that the zeroth order \eqref{eqn:eqnpenroselimit}, with $S^{(0)}(U)$ depending only on $U$, is indeed a solution. In addition, observe that at order $k$, the number of different functions of $U$ in the ansatz, $a^{(k)}_{S,n_v, n_i}(U)$, is given by
\begin{equation}
\#\left(a^{(k)}_{S,n_v, n_i}(U)\right) =\begin{cases}
(1+r)^2  \quad \quad  &\text{even:} \quad $k = 2r$ \\
(2+r)(1+r)  \quad \quad  &\text{odd: } \quad $k = 2r+1$ \\
\end{cases}  \, .
\end{equation} 
In particular, there are two functions $a^{(1)}_{S,n_v, n_i}(U)$ at order 1 and  four functions $a^{(2)}_{S,n_v, n_i}(U)$ at order 2.

\paragraph{First subleading order} The two first order functions $a^{(1)}_{S,n_v, n_i}(U)$ are unconstrained by \eqref{eqn:Seqnspert} as polynomials of degree one in $X^1$ and $X^2$, with coefficients depending only on $U$, are homogeneous solutions to the wave equation on the Penrose limit plane wave. Therefore, \eqref{eqn:Seqnspertexpl} cannot be satisfied by a solution of the form \eqref{eqn:Sexpansion} at first order unless there is no source, $J_S^{(1)} = 0$. Yet the source term at first subleading order is given by
\begin{equation}\label{eqn:source1}
J_S^{(1)} =  -\square^{(1)}S^{(0)},
\end{equation}
where
\begin{equation}
\begin{aligned}
\square^{(1)}  &=  X^i\left(\partial_U   \left(\frac{\left(\partial_i C\right)|_{\gamma}}{2C_{(\gamma)}}\right)\right)  \partial_V -\frac{C^{kj}_{(\gamma)} \left(\partial_k C\right)|_{\gamma}}{2C_{(\gamma)}}  \partial_j -   X^k (\partial_k C^{ij})|_{\gamma}\partial_i\partial_j   -   (\partial_k C^{kj})|_{\gamma}\partial_j  \\
&+\frac{1}{\sqrt{C_{(\gamma)}}}\partial_U B^i_{(\gamma)} \sqrt{C_{(\gamma)}} \partial_i + B_{(\gamma)}^i \partial_i  \partial_U \, .
\end{aligned}
\end{equation}
In \eqref{eqn:source1}, we have used that $\left(\Psi_{ABCD} \Psi^{ABCD}\right)^{1/2} \sim O(\epsilon^2)$. However, from $S^{(0)} = S^{(0)}(U)$, it now follows that indeed $J_S^{(1)} = 0$, as desired for the consistency of the ansatz \eqref{eqn:Seqnspertexpl}.

For the Maxwell equations \eqref{eqn:feqnspert1}-\eqref{eqn:feqnspert4}, we first verify that the leading order of the ansatz \eqref{eqn:fexpansion}, \eqref{eqn:eqnpenroselimit}, is in fact a (source-free) solution. Next, contrary to the scalar case, \eqref{eqn:feqnspert1} does provide a constraint for the ansatz \eqref{eqn:fexpansion}; it is not automatically a source-free solution. On the other hand,   \eqref{eqn:feqnspert2}-\eqref{eqn:feqnspert4} are automatically satisfied at first order by the ansatz \eqref{eqn:fexpansion}, unless they have a source term in which case they are inconsistent (in the sense that they cannot be satisfied). However, with \eqref{eqn:eqnpenroselimit} as the zeroth order input in addition to the geometrical data from the metric \eqref{eqn:adapted} and the frame, we find that the sources for \eqref{eqn:feqnspert2}-\eqref{eqn:feqnspert4} indeed vanish at first order while $J^{(1)}_{f,1} = J^{(1)}_{f,1}(U)$ is just a function of $U$. The equation \eqref{eqn:feqnspert1} will thus generically fix one of the three free functions of $U$ in $f^{(1)}_I$.  \\

Finally, consider the algebraic relations \eqref{eqn:scalarsdcrelations} at first subleading order. These equations can be used to fully fix the Maxwell scalars in terms of the geometric data in addition to $S^{(1)}$ and $S^{(0)}$
\begin{equation}\label{eqn:scalarsdcrelations1}
\begin{aligned}
f_0^{(1)} &= \frac{1}{ 2f_0^{(0)}}(S^{(0)}\Psi_0^{(1)}+\Psi_0^{(0)} S^{(1)})  \, , \\
f^{(1)}_1 &= \frac{S^{(0)}}{f_0^{(0)}} \Psi_1^{(1)} \, , \quad f_2^{(1)} = 0 \, ,
\end{aligned}
\end{equation}
which is just a special case of \eqref{eqn:fexpansiongeneral}. Moreover, from the Penrose expansions of the metric \eqref{eqn:adapted} and the tetrad \eqref{eqn:adaptedframe},  we can find the following structure for the Weyl scalars 
\begin{equation}
\Psi_1^{(1)}  = a^{(1)}_{\Psi_1,0, 0,0}(U) \, , \quad \Psi_0^{(1)} =  a^{(1)}_{\Psi_1,0, 1,0}(U)  X^1 + a^{(1)}_{\Psi_1,0, 0,1}(U)  X^2 \, ,
\end{equation}
also consistent with \eqref{eqn:Psiexpansion}.

If we now try to impose \eqref{eqn:scalarsdcrelations1} together with the first order equations of motion for $S^{(1)}$ and $f^{(1)}_I$, we conclude from our discussion of these equations individually that only \eqref{eqn:feqnspert1} reduces to a non-trivial ordinary differential equation in $U$ but that, inserting \eqref{eqn:scalarsdcrelations1}, we have the two free functions of $U$ in $S^{(1)}$ to satisfy it. Therefore we are able to solve all the Weyl double copy equations but, at this order, these equations do not yet fully fix $S^{(1)}$ and, as a consequence, $f_0^{(1)}$. 

In conclusion, the double copy relations to first order in the Penrose expansion can always be satisfied. 

\paragraph{Second subleading order} At the next order, for the scalar equation of motion \eqref{eqn:Seqnspert}, we find a (generically) non-vanishing source term of the form
\begin{equation}
J_S^{(2)} =  -\square^{(2)}S^{(0)} + \square^{(1)}S^{(1)}+ \left[\left(\Psi_{ABCD} \Psi^{ABCD}\right)^{1/2}\right]^{(2)} \, .
\end{equation}
Contrary to the first order, each term potentially contributes. For instance
\begin{equation}
\begin{aligned}
\sqrt{C_{(\gamma)}} &\square^{(2)}S^{(0)}(U) = \\ & \left\lbrace  \left(\frac{\left(\partial_V C\right)|_{\gamma}}{2\sqrt{C_{(\gamma)}}}\right) +\left. \left(\partial_i(B^i \sqrt{C})\right)\right|_{\gamma}  -  \partial_U \left(D_{(\gamma)} + B^{(\gamma)}_i C^{ij}_{(\gamma)} B^{((\gamma))}_j \right) \sqrt{C_{(\gamma)}} \right\rbrace \partial_U S^{(0)}(U) \, .
\end{aligned}
\end{equation}
Nevertheless, say using \eqref{eqn:Seqnspertexpl} at zeroth and first order, we find a (generically) non-zero source which is just a function of $U$; $J_s^{(2)} = J_s^{(2)}(U)$. From the ansatz \eqref{eqn:Sexpansion}, $S^{(2)}$ is a sum of polynomials of degree one in $V$ and degree two in $X^i$ with $U$-dependent coefficients. Acting with the zeroth order Laplacian on a function of this form simply yields a function of $U$. Therefore, of the four free functions of $U$ in $S^{(2)}$, one will be fixed by \eqref{eqn:Seqnspert} at second order. 

For the single copy equations of motion, although more algebraically cumbersome to make explicit, the equations \eqref{eqn:feqnspert1}-\eqref{eqn:feqnspert3} yield constraints at second order. However, the Maxwell equation \eqref{eqn:feqnspert1} is now linear in the transverse coordinates $X^1$ and $X^2$ and the free functions of $U$ in \eqref{eqn:fexpansion} can only be chosen to make their coefficients vanish separately. Therefore, the three equations \eqref{eqn:feqnspert1}-\eqref{eqn:feqnspert3} impose \emph{four} relations on the seven coefficient functions of  \eqref{eqn:fexpansion} at second order.  The left-hand side of \eqref{eqn:feqnspert4} still vanishes identically on the ansatz \eqref{eqn:fexpansion} and, given \eqref{eqn:fexpansion} at lower orders, so does the source term  $J^{(2)}_{f_4}[f^{0}_I, f^{1}_I]$ on the right-hand side, as required for consistency.

From the algebraic constraints \eqref{eqn:scalarsdcrelations}, we can again fix fully the single copy in terms of the geometry, the zeroth copy, and lower orders as (see \eqref{eqn:fexpansiongeneral})
\begin{equation}\label{eqn:fexpansion2}
\begin{aligned}
f_0^{(2)} &= -  \frac{f_0^{(1)} f_0^{(1)}}{2 f_0^{(0)}} + \frac{1}{ 2f_0^{(0)}}(S^{(0)}\Psi_0^{(2)}+\Psi_0^{(1)} S^{(1)}+\Psi_0^{(0)} S^{(2)})  \, , \\
f^{(2)}_1 &= \frac{S^{(0)}}{f_0^{(0)}} \Psi_1^{(2)} +  \frac{S^{(1)}}{f_0^{(0)}} \Psi_1^{(1)}  - \frac{S^{(0)}}{(f_0^{(0)})^2} f_0^{(1)} \Psi_1^{(1)}  \, , \\  
f_2^{(2)} &= \frac{1}{f_0^{(0)}}\left(S^{(0)}\Psi_2^{(2)}-2 (f_1^{(1)})^2\right) \, .
\end{aligned}
\end{equation}
It is a good consistency check to verify that, for zeroth and double copy of the form \eqref{eqn:Sexpansion} and \eqref{eqn:Psiexpansion} respectively, \eqref{eqn:fexpansion2} yields on single copy of the form  \eqref{eqn:fexpansion}. 

To combine all double copy constraints at second order, it is important to observe that $f^{(2)}_1$ and $f_2^{(2)}$ in \eqref{eqn:fexpansion2} are fully fixed in terms of just the geometry and lower order quantities. Specifically, they do not depend on $S^{(2)}$. Therefore inserting \eqref{eqn:fexpansion2} into \eqref{eqn:feqnspert2}, we find that this equation is just a single function of $U$, as expected, but we \emph{cannot} use one of the three free functions from $S^{(2)}$ to ensure it is satisfied. Nevertheless, one of the two free functions in $S^{(1)}$ was still unspecified and can thus in principle be used instead, leaving us with two constraints from \eqref{eqn:feqnspert1} and one from \eqref{eqn:feqnspert3}, where the three free functions of $U$ from $S^{(2)}$ do feature. Specifically, $\delta^{(0)}f_0^{(2)}$ will leave us able to solve \eqref{eqn:feqnspert3} using $a^{(2)}_{S,1, 0,0}(U)$.  The remaining $a^{(2)}_{S,0, 2,0}(U)$, $a^{(2)}_{S,0, 1,1}(U)$, and $a^{(2)}_{S,0, 0,2}(U)$ should then enable us to solve the two equations from \eqref{eqn:feqnspert1} together with the second order scalar wave equation. 

At the level of counting free functions and equations, we would conclude from the previous that arbitrary spacetimes have a double copy at second order in the Penrose expansion. However, the argument hinges on the existence of solutions to linear systems of the form
\begin{equation}\label{eqn:Meqns}
\begin{aligned}
M^{(1)}(U)\begin{pmatrix}
a^{(1)}_{S,0, 1,0}(U) \\ a^{(1)}_{S,0, 0,1}(U) 
\end{pmatrix} = T^{(1)}(U)  \, , \quad	M^{(2)}(U)\begin{pmatrix}
a^{(2)}_{S,0, 2,0}(U) \\ a^{(2)}_{S,0, 1,1}(U) \\  a^{(2)}_{S,0, 0,2}(U)
\end{pmatrix} = T^{(2)}(U) \, .
\end{aligned}
\end{equation}
where $M^{(1)}(U)$ is a  $2\times 2$ matrix, $M^{(2)}(U)$ a  $3\times 3$ matrix, and $T^{(1)}(U)$ and $T^{(2)}(U)$ are two-and three dimensional vectors. The naive counting argument will go through identically if $M^{(1)}(U)$ and $M^{(2)}(U)$ are invertible. If they are not, the conclusion fails; there is no double copy at second order if any of the sources ($T^{(1)}(U)$) or $T^{(2)}(U)$) falls outside the range of $M^{(1)}(U)$ or  $M^{(2)}(U)$. Even if the sources do fall inside the range, such that the equations can still be solved, the counting argument fails since there is residual freedom (associated to the kernels of $M^{(1)}(U)$ or $M^{(2)}(U)$), that a naive counting of free functions and equations does not account for.

Consider first $M^{(1)}(U)$, which can be read-off from \eqref{eqn:feqnspert1} at first order and \eqref{eqn:feqnspert2} at second order. The only freedoms left in these equations when imposing the double copy relations \eqref{eqn:fexpansiongeneral} and the Penrose expansion ansatz \eqref{eqn:Sexpansion} are $a^{(1)}_{S,0, 1,0}(U)$ and $a^{(1)}_{S,0, 0,1}(U)$. We find
\begin{equation}\label{eqn:M1matrix}
M^{(1)}(U) = \begin{pmatrix}
\frac{\Psi^{(0)}_0}{2\sqrt{2} f_0^{(0)}} \left(E^1_{(1)}+i E^1_{(2)}\right) & \frac{\Psi^{(0)}_0}{2\sqrt{2} f_0^{(0)}} \left(E^2_{(1)}+i E^2_{(2)}\right) \\
\frac{3\Psi^{(1)}_1}{2\sqrt{2} f_0^{(0)}}\left(E^1_{(1)}+i E^1_{(2)}\right) & \frac{3\Psi^{(1)}_1}{2\sqrt{2} f_0^{(0)}} \left(E^2_{(1)}+i E^2_{(2)}\right) 
\end{pmatrix} \, ,
\end{equation}
where the $E^i_{(a)}$ should be evaluated at $\gamma$ ($V=X^i=0$) but we have dropped the subscripts ${}_{(\gamma)}$ (as used in \eqref{eqn:Seqnspertexpl}) here for notational simplicity. In addition, we have used the convention (as in \cite{Chawla:2024mse}) that
\begin{equation}
\bar{m} = -\frac{1}{\sqrt{2}}\left(E_{(1)}+i E_{(2)}\right)  \, .
\end{equation}
Clearly, $M^{(1)}(U)$ is not invertible. \\ 

Similarly, from the field equations \eqref{eqn:Seqnspertexpl} and \eqref{eqn:feqnspert1} (together with \eqref{eqn:fexpansion2}) we read-off
\begin{equation}\label{eqn:Mmatrix}
M^{(2)} = \begin{pmatrix}
2 C^{11} & 2 C^{12} & 2 C^{22} \\
\frac{\Psi^{(0)}_0}{\sqrt{2} f_0^{(0)}} \left(E^1_{(1)}+i E^1_{(2)}\right) & \frac{\Psi^{(0)}_0}{2\sqrt{2} f_0^{(0)}} \left(E^2_{(1)}+i E^2_{(2)}\right) & 0\\
0 & \frac{\Psi^{(0)}_0}{2\sqrt{2} f_0^{(0)}}  \left(E^1_{(1)}+i E^1_{(2)}\right) &  	\frac{\Psi^{(0)}_0}{\sqrt{2} f_0^{(0)}}\left(E^2_{(1)}+i E^2_{(2)}\right)
\end{pmatrix} \, ,
\end{equation} 
where $C^{ij}$ and $E^i_{(a)}$ should again be evaluated at $\gamma$ ($V=X^i=0$). As a starting observation regarding the invertibility of \eqref{eqn:Mmatrix}, we can at least make $C^{ij}_{(\gamma)}(U)$ diagonal at one instant of time, say $U_0$ (a more general attempt at diagonalization would not preserve the form \eqref{eqn:adapted}). For either this $U_0$ or the special case of a diagonalizable Penrose limit \cite{Tod:2019urw} such that we can take  $C^{12} = E^1_{(2)} = E^2_{(1)} = 0$, recalling also the frame \eqref{eqn:adaptedframetransverse}, we find
\begin{equation}\label{eqn:detMdiag}
\begin{aligned}
\text{det}(M_{\rm diag})
&= \frac{1}{2}\left( -C^{11}\left(E^2_{(2)}\right)^2 + C^{22}\left(E^1_{(1)}\right)^2\right) = 0 \, .
\end{aligned}
\end{equation}
In this special case, one can straightforwardly see that the range of $M$ will be two-dimensional. We have verified that, more generally, $\text{det}(M^{(2)}) = 0$ and that the range of $M^{(2)}$ is two dimensional\footnote{Multiply the second row in \eqref{eqn:Mmatrix} by $E^1_{(1)}-i E^1_{(2)}$ and the third by $E^2_{(1)}-i E^2_{(2)}$, and sum them to see that this sum is proportional to the first row using \eqref{eqn:adaptedframetransverse}.}.

We have found that both $M^{(1)}$ and $M^{(2)}$ are not invertible. The result is that the specific forms of $T^{(1)}(U)$ and $T^{(2)}(U)$ will be important in order to decide if the double copy breaks down at second order in the Penrose expansion, and, if it does not, whether there is more residual freedom of $U$ left in  $S^{(1)}$ and $S^{(2)}$.

 The explicit expressions for $T^{(1)}(U)$ and $T^{(2)}(U)$, which we do not display as they are lengthy and not insightful, do not give indication that they should generically be in the range of $M^{(1)}$ and $M^{(2)}$. For instance, in order to be able to satisfy all the double copy relations at second order, the ratio of the two components of  $T^{(1)}(U)$ should equal the ratio of the two rows in $M^{(1)}$, which is $\Psi_0^{(0)}/3\Psi_1^{(1)}$. It is straightforward though tedious, at least without imposing Einstein's equations,  to check that this equality of ratios not in fact satisfied using simple polynomial functions in \eqref{eqn:adapted} with generic coefficients. Therefore, we expect the double copy in the Penrose expansion to generically break down at second subleading order.

\paragraph{Third subleading order}

At third subleading order, while we no longer write out \eqref{eqn:fexpansiongeneral} explicitly, it is still true that only $f_0^{(3)}$ contains a contribution of $S^{(3)}$, $f^{(3)}_1$ and $f^{(3)}_2$ being fixed by lower orders. The same will be true at all orders, only $f_0^{(k)}$ contains $S^{(k)}$. Similarly, while $S^{(k-1)}$ does appear in $f_1^{(k)}$, it does not yet in $f_2^{(k)}$.  As an important consequence, we can only satisfy \eqref{eqn:feqnspert2} and \eqref{eqn:feqnspert4}, which do not involve $f_0^{(k)}$, at a given order by using freedom left from a previous order.

At third order, with the double copy relations satisfied at lower orders, we generically have two degrees of freedom left, one in $S^{(1)}$ and one in $S^{(2)}$, related to the kernels of the matrices $M^{(1)}$ and $M^{(2)}$ defined in \eqref{eqn:M1matrix} and \eqref{eqn:Mmatrix}. Now \eqref{eqn:feqnspert2} constitutes two equations for functions of $U$ as coefficients to a linear function in $X^1$ and $X^2$, as was the case for \eqref{eqn:feqnspert1} at the previous order. However, \eqref{eqn:feqnspert4} is also non-trivial at third order and, based on the left-hand side, does not involve the remaining free function from $S^{(2)}$, but rather only $a^{(2)}_{S,2, 0,0}(U)$, which is not involved in the previously mentioned remaining freedom, having been fixed simply by \eqref{eqn:feqnspert3}. In conclusion, further significant, non-trivial cancellations would be required in order for the double relations to be satisfiable at third order, even at the naive counting level, which, based on the second order discussion, will fail leading to additional constraints.  

As a novelty at third order, we have for the first time an additional constraint from \eqref{eqn:scalarsdcrelations}, as opposed to just fixing $f^{(k)}_I$ in terms of $S^{(k)}_I$, $\Psi^{(k)}_I$, and lower orders:
\begin{equation}\label{eqn:algconstraint3}
S^{(0)}_0 \Psi_3^{(3)} = f^{(1)}_1 f^{(2)}_2 \, .
\end{equation}
We do not have any freedom in the choice of zeroth copy in order to satisfy \eqref{eqn:algconstraint3} with 
\begin{equation}
f^{(1)}_1 = \frac{S^{(0)}}{f_0^{(0)}} \Psi_1^{(1)}  \, , \quad  f^{(2)}_2 = \frac{1}{f_0^{(0)}}\left(S^{(0)}\Psi_2^{(2)}-2 (f_1^{(1)})^2\right) \, .
\end{equation}
In fact, the relation \eqref{eqn:algconstraint3} seems to present a clear constraint on the underlying double copy geometry as it can be expressed entirely in terms of Weyl scalars as
\begin{equation}\label{eqn:algconstraint3psis}
(\Psi_0^{(0)})^2 \Psi_3^{(3)} + 2 ( \Psi_1^{(1)})^3=   \Psi_0^{(0)} \Psi_1^{(1)} \Psi_2^{(2)}\, .
\end{equation}
On the other hand, the equation \eqref{eqn:algconstraint3psis} is not invariant under frame transformations, not even those of the little group, preserving $k$. Therefore, it may well be that \eqref{eqn:algconstraint3psis} can still be satisfied with an appropriate choice of frame, so it may not actually present an invariant geometric constraint. As a result, \eqref{eqn:algconstraint3psis} does not seem to coincide with a known constraint on the Petrov type of the spacetime.

That being said, once a choice of adapted coordinates is made, we can always choose to fix a frame of the form $\left\lbrace k, n,  E_{(a)} \right\rbrace$ defined by \eqref{eqn:adaptedframetransverse} and \eqref{eqn:adaptedframe}. If we impose this specific form, in particular for $n$ in terms of the choice of adapted coordinates and the transverse frame $E_{(a)}$, as a type of gauge-fixing, the only freedom left is really in rotations of the transverse frame, which cannot be used to put \eqref{eqn:algconstraint3psis} to zero. 

To summarize, we have found that in a Penrose expansion around a stretch of a null geodesic free of conjugate points, the Weyl double copy relations can always be satisfied up to first subleading order while they cannot generically be satisfied at second subleading order. At higher orders, the class of spacetimes that admits at perturbative Weyl double copy is restricted still further, as we have illustrated at third subleading order. 

\subsection{Normal neighborhoods} \label{sec:normalneighborhoods}

So far, we have discussed the Penrose expansion and its double copy for a stretch of the target null geodesic $\gamma$ free of conjugate points. As a result, we could use the adapted coordinates \eqref{eqn:adapted} associated to the embedding of (this stretch of) $\gamma$ into a null geodesic congruence. Now, we show how to extend the double copy results to a full normal neighborhood of the null geodesic, which moreover does not depend on a choice and is therefore more directly related to the underlying geometry. 

In terms solely of contractions of the Riemann tensor along a parallel propagated frame along $\gamma$, the metric to second order in the Penrose expansion is given in Fermi null coordinates $y^{\mu} = (u,v,y^{a}) = (u, y^{\bar{a}})$ by \cite{Blau:2006ar}
\begin{equation}\label{eqn:ferminull2}
\begin{aligned}
ds^2 &= 2du dv-R_{a u b u}x^a x^b \, du^2 - \delta_{ab}dx^a dx^b \\
&+ \left\lbrack +\left(2 R_{u a u v}x^a v + \frac{1}{3} (R_{a u b u;c})  x^a x^b x^c \right) du^2 +\frac{4}{3} R_{u b a c}x^b x^c (du dx^a)\right\rbrack  \\
&+ \left[-R_{uvuv}v^2 (du)^2 
-\frac{4}{3} R_{ubvc} x^b x^c (du dv) -\frac{4}{3} R_{uvac}  v
x^c (du dx^a)\right. \\
& -\frac{4}{3} R_{ubav}x^b v (du dx^a)
-\frac{1}{3} R_{acbd} \ x^c x^d (dx^a dx^b) 
-\frac{2}{3} R_{uauv;c} \ x^a x^- x^c (du)^2 \\
& -\frac{1}{3} R_{uaub;v} \ x^a x^b v (du)^2  
-\frac{1}{4} R_{ubac;d} \ x^b x^c x^d (du dx^a) \\
&\left.+(\frac{1}{3} R_{uaAb} R^A_{\;cud} - \frac{1}{12} R_{uaub;cd})  x^a x^b
x^c x^d (du)^2\right] \, .
\end{aligned}
\end{equation}
 As mentioned, the main drawback of \eqref{eqn:ferminull2} is that, to the best of our knowledge, an all orders form of the type \eqref{eqn:ferminull2}  is not known.

The coordinate transformation from adapted coordinates $x^{\mu} =(U,V,X^{i})$ to Fermi null coordinates  $y^{\mu} = (u,v,y^{a}) = (u, y^{\bar{a}})$ associated to the geodesic $\gamma$, as discussed in \cite{Blau:2006ar}, amounts to a transverse Taylor expansion 
\begin{equation}\label{eqn:tofermicoordinates}
x^{\mu}(y^{\nu}) = x^{\mu}(y^{\nu})|_{\gamma} + (\hat{E}^{\mu}_{(\bar{a}_1)}|_{\gamma})y^{\bar{a}_1}-\sum^{\infty}_{n=2}\frac{(-1)^n}{n!}\left( \Gamma^{\mu}_{(\mu_1 \ldots \mu_n)}\hat{E}^{\mu_1}_{(\bar{a}_1)} \ldots \hat{E}^{\mu_n}_{(\bar{a}_n)}|_{\gamma}\right)y^{\bar{a}_1} \ldots y^{\bar{a}_n}  \, ,
\end{equation} 
with the recursively defined generalized Christoffel symbols given by
\begin{equation}
\Gamma^{\mu}_{(\mu_1 \ldots \mu_n)} = \nabla_{(\mu_1}\Gamma^{\mu}_{\mu_2 \ldots \mu_{n})} \, ,
\end{equation}
where the covariant derivative only acts on the lower indices. Here, we are required to use the (extended\footnote{we refer to $y^a$ and $\hat{E}_{(a)}$ as respectively the transverse coordinates and frame while we refer to $y^{\bar{a}}$ and $\hat{E}^{\mu}_{(\bar{a})}$ as respectively the extended transverse coordinates and frame.}) transverse frame $\hat{E}^{\mu}_{(\bar{a})} = \left\lbrace \hat{n}, \hat{E}_{(a)} \right\rbrace$, parallel propagated on $\gamma$ ($V=X_1=X_2 = 0$). As discussed previously, such a frame can always be obtained from $\left\lbrace n, E_{(a)} \right\rbrace$ by frame rotations which leave $k$ invariant. Provided that the leading order scalings are unchanged, such that the frame can still be interpreted as a frame in the Penrose limit, the full analysis of the previous section goes through. Indeed, the explicit form of the frame \eqref{eqn:adaptedframetransverse}-\eqref{eqn:adaptedframe} was not used for the main argument, and was introduced mainly for definiteness and to be able to perform explicit checks.

For convenience of readers less familiar with \cite{Blau:2006ar}, let us briefly illustrate how \eqref{eqn:tofermicoordinates} works for the leading order Penrose limit. In adapted coordinates $\gamma$ is located at $X^1 = X^2 = V = 0$. As a result, the first term in \eqref{eqn:tofermicoordinates} contributes only to $U(u,v,y^a)$. On the other hand, $\hat{E}^{\mu}_{(\bar{a})}$ have no $U$ components and $\Gamma^{U}_{(\mu_2 \ldots \mu_{n})}$ vanishes at the leading Penrose limit order, as $\Gamma^{U}_{\mu \nu}$ then vanishes identically. Thus, at the order of the Penrose limit, $U(u,v,y^a) = U(u)$. Although not strictly enforced by \eqref{eqn:tofermicoordinates}, we set $U = u$, for instance because the Fermi null coordinate $u$ should be affine along $\gamma$ and thus may as well be fixed to be equal to $U$. 

Next, the sum in \eqref{eqn:tofermicoordinates} also does not contribute to $X^i(u,v,y^a)$, now the (generalized) Christoffel symbols $\Gamma^{i}_{(\mu_1 \ldots \mu_n)}$ do not all vanish identically but their projections onto the extended transverse frame vanish nonetheless. As a result, just from the second term in \eqref{eqn:tofermicoordinates}, one finds $X^i = \hat{E}^i_{(a)} y^a$. Finally, for $V(u,v,y^a)$, the second term in \eqref{eqn:tofermicoordinates} contributes a term $v$ but the $n=2$ term of the sum now also contributes through $\Gamma^{V}_{i j} = \frac{1}{2} \partial_U C_{ij}$. 

In summary, 
\begin{equation}\label{eqn:rosenBrinkmann}
\begin{aligned}
U &= u + O(\epsilon)\, , \\
X^i &= \hat{E}^i_{(a)} y^a + O(\epsilon) \, , \\
V &= v - \frac{1}{4}\left(\partial_U C_{ij}\right) \hat{E}^i_{(a)}\hat{E}^j_{(b)} y^a y^b + O(\epsilon) \, . 
\end{aligned}
\end{equation}
is the coordinate transformation between Rosen and Brinkmann coordinates of plane wave spacetimes, which is thus generalized by \eqref{eqn:tofermicoordinates} to all orders. Crucially the leading Penrose expansion scalings of the coordinates is not modified for $(u,v,y^a)$ as compared to $(U,V,X^i)$.

Given the coordinate and frame invariance of the full set of double copy relations \eqref{eqn:scalarsdcrelations} and field equations, we only need to argue that the coordinate transformation \eqref{eqn:tofermicoordinates} respects the order counting that was the organizing principle of Section \ref{eqn:curvedlocal}. Essentially, before rescaling such that the plane wave coordinates are finite, $(u,v,y^a) \sim (U,V,X^i) \sim (\epsilon^0, \epsilon^2, \epsilon)$, which is true on account of \eqref{eqn:rosenBrinkmann}. 

In conclusion, following the steps outlined in Section \ref{sec:overview}, we have first considered a perturbative Weyl double copy in the Penrose expansion for local stretches of a null geodesic in adapted coordinates and subsequently argued that this local perturbative analysis can be extended to a region around the entire null geodesic in null Fermi coordinates. We find that in the Penrose expansion around an arbitrary null geodesic in an arbitrary spacetime, the Weyl double copy relations can always be satisfied up to first subleading order. However, only in special cases do we expect these relations to be satisfiable at higher order.

\section{Kerr-Schild double copy in the Penrose expansion} \label{sec:kerrschild}

In the previous section, we discussed a version of the Weyl double copy in which the single and zeroth copy satisfy field equations on the curved (double copy) background. Here, we wish to consider the case where they are in fact defined on a flat background spacetime. The main new question is how to identify the flat space. Within the classical double copy, a clear prescription is available in the case of the Kerr-Schild double copy \cite{Monteiro:2014cda}. Therefore, we restrict ourselves to that case here.

\subsection{Background}

A Kerr-Schild metric takes the form
\begin{equation}\label{eqn:KS}
g_{\mu \nu} = \eta_{\mu \nu} + \phi k_{\mu} k_{\nu},
\end{equation}
where $k$ is null with respect to both the curved metric $g$ and the flat metric $\eta$. The statement of the Kerr-Schild double copy is then that, if a Kerr-Schild metric \eqref{eqn:KS} satisfies the Einstein equations, then the gauge field
\begin{equation}
A_{\mu} = \phi k_{\mu} \, ,
\end{equation}
satisfies the Maxwell equations, on the flat space in Kerr-Schild coordinates with metric $\eta_{\mu \nu}$, and the scalar field $\phi(x)$ similarly satisfies the flat space massless scalar wave equation \cite{Monteiro:2014cda}.

An important Kerr-Schild background for our purposes is the plane wave spacetime. The Kerr-Schild form \eqref{eqn:KS} is manifest in Brinkmann (Fermi null) coordinates
\begin{equation}\label{eqn:planewave}
ds^2 = 2du dv - \delta_{ab}dx^a dx^b + H(u,x^a) du^2 \, .
\end{equation}
Moreover, the vector field associated to form $du$ is the four-fold repeated principal null direction $\partial_v$. As we emphasized in \cite{Chawla:2024mse}, while we could use the flat space suggested by \eqref{eqn:planewave} to identify a flat space consistent with the double copy of the Penrose limit proposed there, the double copy fields themselves are not the Kerr-Schild double copy as proposed in \cite{Monteiro:2014cda}. This difference can be seen directly from the fact that $H(u,x^a)$ is a quadratic form in $x^a$, while the Penrose limit expectation is that $\phi(x^{\mu}) = \phi(u) \neq  H(u,x^a)$.

\subsection{Example: Schwarzschild}\label{sec:example:Schwarzschild}

Consider the Schwarzschild metric in Kerr-Schild coordinates
\begin{equation}\label{eqn:SmetricKS}
ds^2 = d\hat{t}^2- dr^2-r^2 \left(d\theta^2 + \sin^2\theta d \phi^2\right) -\frac{2M}{r}(d\hat{t} + dr)^2  \, .
\end{equation}
In adapted coordinates $\lbrace U,V,\hat{t}_0,\hat{\phi}_0 \rbrace$ we find
\begin{equation}\label{eqn:Sadapted}
\begin{aligned}
ds^2 =  
2 dU dV&- \left(r^2 \frac{\hat{E}^2}{L^2}-\left(1-\frac{2M}{r}\right)\right)d\hat{t}_0^2 - r^2 \sin^2(\theta) d \phi_0^2  \\ &-\frac{r^2}{L^2} dV^2+2\hat{E} \frac{r^2}{L^2} dV d\hat{t}_0 \, .\\
\end{aligned}
\end{equation}
where $r = r(U;M)$ and $\theta = \theta(U,V,\hat{t}_0;M)$ are defined as functions of the adapted coordinates in more detail in Appendix \ref{app:Penroseexpansion}. We emphasize especially the dependence of these functions on $M$ (as opposed to the additional dependence on say $L/\hat{E}$) to distinguish $M=0$ versions of these functions if required in the single copy.

In adapted coordinates, the single copy gauge field takes the form
\begin{equation}\label{eqn:AKS}
A = \frac{2M}{r}(d\hat{t} + dr) = \frac{2M}{r}\left(d\hat{t}_0 + \hat{E} dU \left(1+\frac{2 M r + r^2\sqrt{1-\frac{L^2}{\hat{E}^2 r^2}(1-\frac{2M}{r})}}{r(r-2M)}\right) \right) \, .
\end{equation}
Importantly for the Penrose limit, and by extension the Penrose expansion, \eqref{eqn:AKS} is not in Penrose-G\"uven gauge; that is $A_U \neq 0$. Relatedly, going through the adapted coordinate transformation starting from the flat Kerr-Schild metric
\begin{equation}\label{eqn:SmetricKSflat}
ds^2_{\flat} = d\hat{t}^2- dr^2-r^2 \left(d\theta^2 + \sin^2\theta d \phi^2\right)  \, ,
\end{equation} 
we find a metric which is not in the adapted coordinate form; $g_{UU} \neq 0 \neq g_{Ui}$. Having a metric which is not in adapted coordinates is a problem from the perspective of taking the Penrose limit, as the leading order scaling would not be correct; there would be an $O(\epsilon^0)$ term, dropping all higher orders the metric would be degenerate, and the gauge field would be pure gauge. 

Given that $r=r(U)$, we can readily shift the gauge to remove the second term in \eqref{eqn:AKS}. In fact, subsequently taking the Penrose limit for the gauge field yields a field strength
\begin{equation}\label{eqn:SFgamma}
F^{(\gamma)} = -\frac{2M r'(U)}{r^2} dU \wedge d\hat{t}_0 = - \frac{2M L}{r^3} du \wedge dx_1 
\end{equation}
where in the second line we have gone to the Brinkmann coordinates using
\begin{equation}
\hat{t}_0= \frac{L x_1}{r \sqrt{ \hat{E}^2-\frac{L^2}{r^2}(1-\frac{2M}{r})}} \, , \quad r'(U) = r'(u) =  \sqrt{\hat{E}^2-\frac{L^2}{r^2}(1-\frac{2M}{r})} \, .
\end{equation}
The field strength \eqref{eqn:SFgamma} is the result consistent with the Weyl double copy as already discussed in the Penrose limit in \cite{Chawla:2024mse} \footnote{Note that an additional factor of $1/\sqrt{2}$ was included there in the definition of the Kerr-Schild vector $k_{\mu}$.}.

On the other hand, consider how gauge transformations, such as those needed to go to Penrose-G\"uven gauge, affect the Kerr-Schild double copy. After a gauge transformation, the Kerr-Schild form \eqref{eqn:KS} looks like
\begin{equation}\label{eqn:KSgauge}
g_{\mu \nu} = \left\lbrack \eta_{\mu \nu} + \frac{2 A_{(\nu} \partial_{\mu)} \Lambda}{\phi} + \frac{\partial_{\mu} \Lambda \partial_{\nu}\Lambda}{\phi} \right\rbrack + \frac{A_{\mu} A_{\nu}}{\phi}  \, .
\end{equation}
For instance, for Schwarzschild, it is fairly natural to perform a gauge transformation in order to put the single copy into the usual form of the electrostatic potential. That is, instead of \eqref{eqn:SmetricKS}, we write
\begin{equation}\label{eqn:SmetricKSstat}
ds^2 = \left\lbrack d\hat{t}^2- dr^2-r^2 \left(d\theta^2 + \sin^2\theta d \phi^2\right)  -\frac{4M}{r} dr d\hat{t} -\frac{2M}{r}dr^2 \right\rbrack -\frac{2M}{r}d\hat{t}^2  \, .
\end{equation}
This is exactly the same metric, we have just shuffled around the terms and called it a gauge transformation for the single copy but
\begin{equation}\label{eqn:SmetricKSstatnotflat}
ds^2 =  \left\lbrack d\hat{t}^2- dr^2-r^2 \left(d\theta^2 + \sin^2\theta d \phi^2\right)  -\frac{4M}{r} dr d\hat{t} -\frac{2M}{r}dr^2 \right\rbrack \, ,
\end{equation}
is not a flat metric. That the identification of the Kerr-Schild flat coordinates is gauge dependent is well-known, and not a surprise considering amplitude double copy relations. However, as we have illustrated, taking the Penrose limit through adapted coordinates generically involves an incompatible gauge choice.

Let us nevertheless take the Penrose limit up to the usual order ($O(\epsilon)$ for the gauge field and $O(\epsilon^2)$ for the metric) for the flat space metric in adapted coordinates. We find
\begin{equation}\label{eqn:Sadaptedpenroseflat}
\begin{aligned}
ds^2_{\gamma,\flat} =  
2 dU dV&- \left(r^2 \frac{\hat{E}^2}{L^2}-1\right)d\hat{t}_0^2 - r^2 \sin^2 \theta d \phi_0^2 \\ &+ \frac{4 M \hat{E}}{r\left(1-\frac{2M}{r}\right)}\left(1+\sqrt{1-\frac{L^2}{\hat{E}^2 r^2}\left(1-\frac{2M}{r}\right)}\right)dU d\hat{t}_0 \\ &+ \frac{2M \hat{E}^2}{r\left(1-\frac{2M}{r}\right)}  \left(2-\frac{L^2}{\hat{E}^2 r^2}\left(1-\frac{2M}{r}\right)+2\sqrt{1-\frac{L^2}{\hat{E}^2 r^2}\left(1-\frac{2M}{r}\right)}\right) dU^2 \, ,
\end{aligned}
\end{equation}
where we still have $r = r(U;M)$ and $\theta = \theta(U;M)$ depending on $M$ because, despite representing flat space, the mass was introduced into \eqref{eqn:Sadaptedpenroseflat} by following the exact coordinate transformations that were used for the curved metric, including this mass dependence.

Similarly going through the coordinate transformation used for the curved metric to go to Brinkmann coordinates, we find
\begin{equation}\label{eqn:Sadaptedpenroseflatbrinkmann}
\begin{aligned}
ds^2_{\gamma,\flat} &=  
2 du dv -dx^2_1 - dx_2^2\, \\
&+ 2 M x_1  \frac{2 L^2 (L^2 M - \hat{E}r^3) du dx_1}{r^{5/2} \left(2 L^2 M -L^2 r + \hat{E} r^3 \right)^{3/2}}  + \frac{2L^2 M}{2 L^2 M -L^2 r + \hat{E}^2 r^3} dx_1^2 \\
&+ \left( \frac{3 L^2 M x_2^2}{r^5} + \frac{2M \hat{E}^2}{r} X_0(u) + \frac{4 M L \hat{E} x_1}{r^3} X_1(u) + \frac{M L^2 x_1^2}{r^5} X_2(u) \right)du^2 \, ,
\end{aligned}
\end{equation}
where the somewhat lengthy functions $X_0, X_1, X_2$, whose details we don't need, are given in  Appendix \ref{app:Penroseexpansion}. The metric \eqref{eqn:Sadaptedpenroseflatbrinkmann} is clearly not in the Brinkmann form for flat space, exactly because we used the coordinate transformation that brings the \emph{curved} spacetime into Brinkmann form. Of course, we still recover the usual flat space expression by setting $M = 0$. Although these $M = 0$ coordinates are consistent with the Penrose limit Weyl double copy single and zeroth copy satisfying the wave equation on them \cite{Chawla:2024mse}, they are thus not the same as what we find directly by mimicking identically every coordinate transformation needed to find the Penrose limit for the curved spacetime on the Kerr-Schild flat space.

As an intermediate option between letting $M = 0$ in \eqref{eqn:Sadaptedpenroseflatbrinkmann} and taking the full $M \neq 0$ form, it seems natural, from the perspective of having a homogeneously $O(\epsilon^2)$ metric, to take $M \sim \epsilon^2$. With such a scaling, in adapted coordinates
\begin{equation}\label{eqn:Sadaptedpenroseflatscaling}
\begin{aligned}
ds^2_{\gamma,\flat} =  
2 dU dV&- \left(r^2 \frac{\hat{E}^2}{L^2}-1\right)d\hat{t}_0^2 - r^2 \sin^2 \theta d \phi_0^2 \\  &+ \frac{2M \hat{E}^2}{r}  \left(2-\frac{L^2}{\hat{E}^2 r^2}+2\sqrt{1-\frac{L^2}{\hat{E}^2 r^2}}\right) dU^2 \, ,
\end{aligned}
\end{equation}
where now $r = r(U;0)$ and $\theta = \theta(U;0)$, while in Brinkmann coordinates
\begin{equation}\label{eqn:Sadaptedpenroseflatbrinkmannscaling}
\begin{aligned}
ds^2_{\gamma,\flat} &=  
2 du dv -dx^2_1 - dx_2^2 + \frac{2M \hat{E}^2}{r} \left(2- \frac{L^2}{r^2 \hat{E}^2} +2\sqrt{1-\frac{L^2}{\hat{E}^2 r^2}}\right) du^2 \, .
\end{aligned}
\end{equation}
Here, $M$ still appears non-trivially and we still do not find the flat space that would have been expected from the plane wave Kerr-Schild double copy. 

Finally, we could choose $M \sim \epsilon^{2 + \delta}$ with $\delta > 0$. This choice absorbs a homogeneous $O(\epsilon^2)$ scaling of the metric. It does reduce to the natural plane wave flat space, and with the inclusion of all orders should nevertheless reproduce the full flat, Schwarzschild Kerr-Schild coordinates. It is not surprising that, for $M \sim \epsilon^{2 + \delta}$, the expected Kerr-Schild flat space emerges with respect to the leading order plane wave as well as with respect to the full black hole, as when setting $M=0$ identically in the black hole metric \eqref{eqn:SmetricKS} and the Penrose limit in Brinkmann coordinates \eqref{eqn:brinkmann}.

However, if the original Kerr-Schild coordinates are to be identified between the curved and the flat space, the flat space coordinate transformations generically include non-trivially $M$-dependence in order to keep track of that flat space when going to adapted coordinates. Therefore, a prescription $M \sim \epsilon^{2 + \delta}$ must still be used instead of simply setting $M = 0$ identically in order not to lose track of the exact flat space identification. 

Similar considerations related to identifying the appropriate flat space can be made for the asymptotic double copy. In particular, in \cite{Godazgar:2021iae}, both setting the curved space mass, and similar charges, to zero as well as going from Kerr-Schild coordinates to (flat space) Bondi coordinates are discussed. Nevertheless, an important difference is that the coordinate transformation from Kerr-Schild to adapted coordinates around arbitrary null geodesics could naturally depend on the mass in a more intricate way than the analogous transformation from Kerr-Schild to Bondi coordinates. On the other hand, the dependence of the coordinate transformation from the Kerr-Schild to adapted (or Bondi) coordinates is exactly how the prescription $M \sim \epsilon^{2 + \delta}$ will end up differing from setting $M = 0$ identically.

In summary, the Schwarzschild example illustrates that, in trying to follow through what happens step-by-step with the Kerr-Schild flat space coordinates in the Penrose expansion, we are faced with the incompatibility of the Kerr-Schild gauge and the Penrose-G\"uven gauge. Three possible resolutions are (i) to not truncate at the superleading $O(\epsilon^0)$ but nevertheless going to $O(\epsilon^2)$, as usual, and declaring these combined three orders to be leading, (ii) taking $M \sim \epsilon^2$ in the construction of the flat space metric, or (iii) taking $M \sim \epsilon^{2 + \delta}$ with $\delta > 0$. In order to have a viable Penrose expansion (i) seems disfavored while, in order for the leading order to be the Kerr-Schild flat space of the plane wave (iii) seems desirable.

\section{Conclusion and outlook}

We have formulated a Weyl double copy on a curved background in the Penrose expansion. Exact Weyl double copy spacetimes preserve their double copy structure order-by-order in this expansion. On the other hand, we have argued that generic spacetimes, even without exact double copy structure, still have a double copy in the Penrose expansion to first subleading order. 

As a result, we have established a precise new interface to study how the classical Weyl double copy breaks down for generic spacetimes. Our approach, using an expansion around null geodesics, is complementary to the asymptotic double copy, formulated perturbatively around null infinity \cite{Godazgar:2021iae,Adamo:2021dfg}. Advantages of the Penrose expansion compared to such an asymptotic approach include uses in non-asymptotically flat spaces as well as interesting spacetime regions such as light-rings, horizon, or near singularities of black holes to which the asymptotic expansion may not be sensitive.

We have provided evidence that the double copy in the Penrose expansion will generically break down at second order. Nevertheless, a better understanding of the actual obstruction to the double copy at second order as well as higher orders would be desirable. One straightforward way forward could be to study what goes wrong in various examples that are not expected to have an all orders Weyl double copy. On the other hand, as such examples will have more generic Petrov types than type D and type N, the Penrose expansion may rapidly grow in complexity. Moreover, many interesting examples such as the family of STU black holes studied in the asymptotic double copy approach \cite{Godazgar:2021iae} or the FLRW spacetimes in an all-orders ``background field'' double copy \cite{Ilderton:2024oly} require (in principle) a treatment of additional field content or sources \cite{Cardoso:2016amd,LopesCardoso:2018xes,Easson:2021asd,Easson:2022zoh,Armstrong-Williams:2024bog}.

In addition, while we are able to formulate an order-by-order Weyl double copy in the Penrose expansion on a curved background, we have not uniquely identified how to construct a flat spacetime with respect to which we can do the same. We have illustrated how to follow through in principle what is the relevant flat space inherited from an exact Kerr-Schild background order-by-order in the Penrose expansion, analogous to an approach suggested for the asymptotic double copy in \cite{Godazgar:2021iae}. However, this Penrose expanded flat space does not seem to have a clear structure in the Penrose expansion as a result of the (generic) incompatibility between the Kerr-Schild gauge and the Penrose-G\"uven gauge. Therefore, it remains an open problem to understand a constructive way to determine a suitable flat space without knowledge of the all-orders background. 

A surprising feature of the Weyl classical double copy is that it is local in position space while double copy constructions in scattering amplitudes are local in momentum space \cite{Monteiro:2021ztt,Luna:2022dxo}. It would be interesting to also understand  the interplay between the Penrose expansion and versions of the double copy which are not generically local in position space such as the convolutional double copy \cite{Anastasiou:2014qba,Anastasiou:2018rdx,LopesCardoso:2018xes,Luna:2020adi} or the twistor double copy \cite{White:2020sfn,Chacon:2021wbr}. A first step in such a direction would likely need to be a clarification of what the Penrose expansion even means in such formulations and in flat space, momentum space scattering amplitudes themselves. The latter, if it can be made into a well-posed question, would likely involve a combination of techniques such as: obtaining classical spacetimes from amplitudes \cite{Boulware:1968zz,Duff:1973zz,Neill:2013wsa,Mougiakakos:2020laz}, eikonal methods \cite{DiVecchia:2023frv} and perhaps a suggestive relationship between perturbative expansions of plane wave background-field amplitudes and multicollinear limits of vacuum (flat space) amplitudes \cite{Adamo:2021hno}.

Finally, the Penrose limit, Fermi null coordinates, and the Penrose expansion have been studied in various applications and a natural question for future work is relating these to what we have done here. One striking application of the Penrose expansion is in AdS/CFT, where string theory in the Penrose expanded backgrounds are dual to particular large charge expansions in a holographically dual field theory \cite{Berenstein:2002jq,Callan:2003xr,Callan:2004uv,McLoughlin:2004dh,Minahan:2005mx,Minahan:2005qj}. The context is very different from what we have discussed here but it should be worthwhile to connect to this AdS/CFT literature more closely, especially in light of the origins of the double copy in the KLT relations in string theory \cite{Kawai:1985xq}. Moreover, there is at least already a commonality in approach between our work and ways in which the Penrose expansion has been used to study loss of integrability \cite{Beisert:2010jr}, on both sides of the AdS/CFT duality, away from the most studied special limits and backgrounds \cite{Asano:2015qwa,Alencar:2021ljc,McLoughlin:2022jyt}. 

As another example application, the Penrose limit of the light-ring captures properties of eikonal quasinormal modes \cite{Fransen:2023eqj,Kapec:2024lnr}. Higher orders in the Penrose expansion are similarly expected to map onto higher orders of the eikonal expansion. We have shown that the double copy structure of the four-dimensional black holes of general relativity in particular are reflected order-by-order in the Penrose expansion. Therefore, the double copy structure of these black holes should also be reflected order-by-order in their eikonal quasinormal modes spectrum. If so, then perhaps the full spectrum may also reflect the underlying double copy.  Such a structure might be apparent in future gravitational wave experiments  \cite{Bhagwat:2021kwv,Bhagwat:2023jwv,Pitte:2024zbi}.

\acknowledgments

We thank Ricardo Monteiro, Silvia Nagy, and Chris White for insightful discussions. S.~C. and C.~K. are supported by the U.S. Department of Energy under grant number DESC0019470 and by the Heising-Simons Foundation “Observational Signatures of Quantum
Gravity” collaboration grants 2021-2818 and 2024-5305. This research was supported in part by grant NSF PHY-2309135 to the Kavli Institute for Theoretical Physics (KITP); C.~K. appreciates the support and hospitality of KITP during the finishing stages of this project.  C.~K. also acknowledges Queen Mary University and Caltech for hospitality during earlier stages of this work.
K.~F. is supported by the Heising-Simons Foundation ``Observational Signatures of Quantum Gravity'' collaboration grant 2021-2817, the U.S. Department of Energy, Office of Science, Office of High Energy Physics, under Award No. DE-SC0011632, and the Walter Burke Institute for Theoretical Physics. C.~K. and K.~F. acknowledge the Aspen Center for Physics, where part of this work was performed, which is supported by National Science Foundation grant PHY-2210452.

\begin{appendices}

\section{Spinor conventions and Newman-Penrose formalism}\label{app:NP}

Our conventions are identical to those we used in \cite{Chawla:2024mse} but we repeat them here to be
self-contained. We borrow our spinor conventions predominantly from \cite{Penrose:1985bww} and \cite{Stephani:2003tm}. Wherever the
two conflict, we prefer conventions from \cite{Penrose:1985bww}. We work in mostly-minus signature $(+\ -\ -\ -)$, and we define the spinor antisymmetric product as
\begin{equation} \label{} 
\kappa_{A} \epsilon^{AB} \tau_{B} = \kappa_{A} \tau^{A} = \kappa^{A} \epsilon_{AB} \tau^{B}.
\end{equation}
This also encapsulates our convention for raising and lowering spinor indices. Explicitly, we use
\begin{equation} \label{} 
\kappa^{A} = \epsilon^{AB} \kappa_{B}, \qquad \kappa_{A} = \kappa^{B} \epsilon_{BA}.
\end{equation}

The only higher-rank tensors we are concerned with are the Weyl tensor $C_{\alpha \beta \gamma \delta}$ and the Maxwell
field strength $F_{\alpha \beta}$. Their spinor counterparts can be expressed using the abstract index notation of
\cite{Penrose:1985bww} as
\begin{align} \label{} 
C_{\alpha \beta \gamma \delta} &= \Psi_{ABCD} \overline{\epsilon}_{A' B'} \overline{\epsilon}_{C' D'} + \text{c.c.}, \\
F_{\alpha \beta} &= f_{AB} \overline{\epsilon}_{A'B'} + \text{c.c.},
\end{align}
where ``c.c.'' is the complex conjugate of the preceding expression, and $f_{AB}$ as well as $\Psi_{ABCD}$ are
fully symmetric in their indices with no other constraints. Thus, the algebraically independent components of the
Maxwell and Weyl tensors are reorganized into fully symmetric spinor objects with no remaining algebraic
constraints.

We use labels $(k, n, m, \overline{m})$ for the Newman-Penrose tetrad, where for a real spacetime
$k$ and $n$ are real, while $m$ and $\overline{m}$ are complex. The tetrad
members are all null, with a relative normalization given by
\begin{equation} \label{} 
g_{\alpha \beta} k^{\alpha} n^{\beta} = 1, \qquad g_{\alpha \beta} m^{\alpha} \overline{m}^{\beta} = -1.
\end{equation}
We have also used the real orthonormal spacelike vectors $E_{(a)}$ ($a = 1, 2$) in terms of which the
complex null vectors are given by
\begin{equation} \label{} 
m = -\frac{1}{\sqrt{2}} \left(E_{(1)} - i E_{(2)}\right), \qquad
\overline{m} = -\frac{1}{\sqrt{2}} \left(E_{(1)} + i E_{(2)}\right).
\end{equation}

We use a spinor dyad given by $(o, \iota )$, which in the abstract index notation is
related to the Newman-Penrose tetrad via
\begin{equation} \label{} 
k_{\alpha} = o_{A} \overline{o}_{A'}, n_{\alpha} = \iota_{A} \overline{\iota}_{A'}, m_{\alpha} = o_{A}
\iota_{A'}, \overline{m}_{\alpha} = \iota_{A} \overline{o}_{A'}.
\end{equation}
The normalization of the Newman-Penrose tetrad induces the spinor dyad normalization
\begin{equation} \label{} 
o_{A}\iota^{A} = 1 = -\iota_{A}o^{A}.
\end{equation}
We will henceforth call this spinor dyad the Newman-Penrose dyad. In this dyad, the spinors $f_{AB}$ and
$\Psi_{ABCD}$ have components $f_{i}$ and $\Psi_{i}$ defined by
\begin{align} \label{} 
f_{AB} &= f_{0} \iota_{A} \iota_{B} - 2f_{1} \iota_{(A} o_{B)} + f_{2} o_{A} o_{B}, \\
\Psi_{ABCD} &= \Psi_{0} \iota_{A} \iota_{B} \iota_{C} \iota_{D} - 4 \Psi_{1} \iota_{(A} \iota_{B} \iota_{C}
o_{D)} + 6 \Psi_{2} \iota_{(A} \iota_{B} o_{C} o_{D)} \\
&\phantom{=\ } - 4 \Psi_{3} \iota_{(A} o_{B} o_{C} o_{D)} + \Psi_{4}
o_{A} o_{B} o_{C} o_{D}.
\end{align}
The spinor components $f_{i}$ and $\Psi_{i}$ can be found using the corresponding tensors via
\begin{subequations}
	\begin{align} 
	f_{0} &= F_{\alpha \beta} k^{\alpha} m^{\beta}, \\
	f_{1} &= \frac{1}{2} F_{\alpha \beta} \left( k^{\alpha} n^{\beta} - m^{\alpha} \overline{m}^{\beta}\right)
	,\\
	f_{2} &= F_{\alpha \beta} \overline{m}^{\alpha} n^{\beta}.
	\end{align}
	\text{and}
	\begin{align} 
	\Psi_{0} &= -C_{\alpha \beta \gamma \delta} k^{\alpha} m^{\beta} k^{\gamma} m^{\delta} \\
	\Psi_{1} &= -C_{\alpha \beta \gamma \delta} k^{\alpha} n^{\beta} k^{\gamma} m^{\delta} \\
	\Psi_{2} &= -C_{\alpha \beta \gamma \delta} k^{\alpha} m^{\beta} \overline{m}^{\gamma} n^{\delta} \\
	\Psi_{3} &= -C_{\alpha \beta \gamma \delta} k^{\alpha} n^{\beta} \overline{m}^{\gamma} n^{\delta} \\
	\Psi_{4} &= -C_{\alpha \beta \gamma \delta} n^{\alpha} \overline{m}^{\beta} n^{\gamma}
	\overline{m}^{\delta}.
	\end{align}
\end{subequations}

Covariant derivatives of spinors can be written in terms of spin coefficients
\begin{equation}\label{eqn:spinorspincoeffs}
\gamma_{AA' C}{}^{B} = \epsilon_{A}{}^B \nabla_{AA'}\epsilon_{C}{}^A \, ,
\end{equation}
such that
\begin{equation}
\begin{aligned}
\nabla_{\mu}\kappa^A &= \nabla_{\mu}\left(\kappa^o o^A + \kappa^{\iota}\iota^A\right) \\
&= \left(\partial_{\mu}\kappa^o  + \gamma_{\mu o}{}^{o}\kappa^o + \gamma_{\mu \iota}{}^{o}\kappa^{\iota}\right) o^A + \left(\partial_{\mu} \kappa^{\iota}+ \gamma_{\mu o}{}^{\iota}\kappa^o + \gamma_{\mu \iota}{}^{\iota}\kappa^{\iota} \right) \iota^A \, .
\end{aligned}
\end{equation}
To compute \eqref{eqn:spinorspincoeffs}, we will use the associated Newman-Penrose frame. Traditionally, names are
given to the individual components \cite{Penrose:1985bww}
\begin{alignat}{5}
\kappa &= m^{\alpha} k^{\mu} \nabla_{\mu} k_{\alpha} \, , \quad \epsilon&=& \frac{1}{2}\left(n^{\alpha} k^{\mu} \nabla_{\mu} k_{\alpha} + m^{\alpha} k^{\mu} \nabla_{\mu} \bar{m}_{\alpha}\right) \, , \quad &\gamma'&= \frac{1}{2}\left(k^{\alpha} k^{\mu} \nabla_{\mu} n_{\alpha} + \bar{m}^{\alpha} k^{\mu} \nabla_{\mu} m_{\alpha}\right) \, , \nonumber \\
\rho &= m^{\alpha} \bar{m}^{\mu} \nabla_{\mu} k_{\alpha} \, , \quad \alpha &=& \frac{1}{2}\left(n^{\alpha} \bar{m}^{\mu} \nabla_{\mu} k_{\alpha} + m^{\alpha} \bar{m}^{\mu} \nabla_{\mu} \bar{m}_{\alpha}\right) \, , \quad &\beta'&  = \frac{1}{2}\left(k^{\alpha} \bar{m}^{\mu} \nabla_{\mu} n_{\alpha} + \bar{m}^{\alpha} \bar{m}^{\mu} \nabla_{\mu} m_{\alpha}\right) \, , \nonumber \\
\sigma &= m^{\alpha} m^{\mu} \nabla_{\mu} k_{\alpha} \, , \quad \beta &=& \frac{1}{2}\left(n^{\alpha} m^{\mu} \nabla_{\mu} k_{\alpha} + m^{\alpha} m^{\mu} \nabla_{\mu} \bar{m}_{\alpha}\right) \, , \quad &\alpha'&= \frac{1}{2}\left(k^{\alpha} m^{\mu} \nabla_{\mu} n_{\alpha} + \bar{m}^{\alpha} m^{\mu} \nabla_{\mu} m_{\alpha}\right) \, , \nonumber \\
\tau &= m^{\alpha} n^{\mu} \nabla_{\mu} k_{\alpha} \, , \quad \gamma&=& \frac{1}{2}\left(n^{\alpha} n^{\mu} \nabla_{\mu} k_{\alpha} + m^{\alpha} n^{\mu} \nabla_{\mu} \bar{m}_{\alpha}\right) \, , \quad &\epsilon'&= \frac{1}{2}\left(k^{\alpha} n^{\mu} \nabla_{\mu} n_{\alpha} + \bar{m}^{\alpha} n^{\mu} \nabla_{\mu} m_{\alpha}\right) \, , \nonumber \\
\tau' &= \bar{m}^{\alpha} k^{\mu} \nabla_{\mu} n_{\alpha} \, , \quad \sigma' &=& \bar{m}^{\alpha} \bar{m}^{\mu} \nabla_{\mu} n_{\alpha} \, , \quad \rho'  = \bar{m}^{\alpha} m^{\mu} \nabla_{\mu} n_{\alpha} \, , \quad &\kappa'&  =\bar{m}^{\alpha} n^{\mu} \nabla_{\mu} n_{\alpha} \, . \label{eqn:spincoeffssymbols}
\end{alignat}
We will also use the following notation to abbreviate directional derivatives in the Newman-Penrose basis,
\begin{align} \label{eqn:directionalderivatives} 
D &\coloneqq k^{\alpha} \nabla_{\alpha} = -o^{A} \overline{o}^{B'} \nabla_{AB'}, \\
\Delta &\coloneqq n^{\alpha} \nabla_{\alpha} = - \iota^{A} \overline{\iota}^{B'} \nabla_{AB'}, \\
\delta &\coloneqq m^{\alpha} \nabla_{\alpha} = -o^{A} \overline{\iota}^{B'} \nabla_{AB'}, \\
\overline{\delta} &\coloneqq \overline{m}^{\alpha} \nabla_{\alpha} = -\iota^{A} \overline{o}^{B'} \nabla_{AB'}.
\end{align}
Using the above definitions of spin coefficients and the directional derivatives, the Maxwell equations,
\begin{equation} \label{} 
\nabla_{AB'} f^{AB} = 0,
\end{equation}
can be expanded in the Newman-Penrose dyad and expressed as
\begin{align}
D f_{1} - \overline{\delta} f_{0} &= -\left(-\tau' + 2 \beta'\right) f_{0} - 2 \rho f_{1} + \kappa f_{2}, \\
D f_{2} - \overline{\delta} f_{1} &= -\sigma' f_{0} + 2 \tau' f_{1} - \left(\rho - 2 \varepsilon\right) f_{2}, \\
\delta f_{1} - \Delta f_{0} &= -\left(-\rho' + 2 \varepsilon' \right) f_{0} - 2 \tau f_{1} + \sigma f_{2}, \\
\delta f_{2} - \Delta f_{1} &= -\kappa' f_{0}  + 2 \rho' f_{1} - \left(\tau - 2 \beta\right) f_{2}.
\end{align}
This general Newman-Penrose form is the starting point for the perturbative Penrose expansion \eqref{eqn:feqnspert1}-\eqref{eqn:feqnspert2} in the main text.

\section{The Penrose expansion}\label{app:Penroseexpansion}

We discuss in this appendix the leading order scaling of various quantities when subject to the Penrose expansion
described in section \ref{eqn:curvedlocal}. The adapted coordinates $\left(U, V, X^{i}\right)$ are substituted by
scaled versions $\left(\epsilon^{0} U, \epsilon^{2} V, \epsilon^{1} X^{i}\right)$, which results in the scaled
metric in adapted coordinates,
\begin{equation} \label{eqn:scaledmetric} 
ds^2 = 2 \epsilon^{2} dU dV + \epsilon^{4} D(x^{\mu}) dV^2 + 2 \epsilon^{3} B_i(x^{\mu}) dV dX^i - \epsilon^{2}
C_{ij}(x^{\mu})dX^i dX^j \ .
\end{equation}
The coordinate scaling implies that the Newman-Penrose tetrad from \eqref{eqn:adaptedframetransverse} and
\eqref{eqn:adaptedframe} scales as
\begin{equation} \label{} 
k^{\mu} \partial_{\mu} = \partial_{U}, n^{\mu} \partial_{\mu} = \epsilon^{-2} \partial_{V} + \epsilon^{-1}
B^{(a)} E_{(a)} - \frac{1}{2} \left( B_{(a)} B^{(a)} + D\right) \partial_{U}, E_{(a)} = \epsilon^{-1}
E^{i}_{(a)} \partial_{i}.
\end{equation}
Thus the leading order scaling is $\left\lbrace k^{\mu}, n^{\mu},  E^{\mu}_{(a)} \right\rbrace \sim \left\lbrace
\epsilon^0, \epsilon^{-2},  \epsilon^{-1} \right\rbrace $. The leading order scaling of the metric itself is
$\epsilon^{2}$, thus
\begin{align} \label{} 
k_{\mu} &= g_{\mu \nu} k^{\nu} \sim \epsilon^{2} \\
n_{\mu} &= g_{\mu \nu} n^{\nu} \sim \epsilon^{0} \\
E_{\mu}^{i} &= g_{\mu \nu} E^{\nu}_{i} \sim \epsilon^{1}.
\end{align}

The leading order scaling of the Newman-Penrose spin coefficients computed using \eqref{eqn:spincoeffssymbols} can now be read
off using the scalings of the tetrad computed above. Explicitly, we have
\begin{alignat}{4} 
\kappa& \sim \epsilon^{0} \quad & \sigma& \sim \epsilon^{0} \quad & \epsilon& \sim \epsilon^{0} \quad & \rho&
\sim \epsilon^{0} \nonumber \\
\tau'& \sim \epsilon^{1} \quad & \tau& \sim \epsilon^{1} \quad & \alpha& \sim \epsilon^{1} \quad & \beta& \sim
\epsilon^{1} \nonumber \\
\sigma'& \sim \epsilon^{2} \quad & \epsilon'& \sim \epsilon^{2} \quad & \rho'& \sim \epsilon^{2} \quad &
\kappa'& \sim \epsilon^{3} \nonumber \\
\alpha'& \sim \epsilon^{1} \quad & \beta'& \sim \epsilon^{1} \quad & \gamma& \sim \epsilon^{2} \quad & \gamma'&
\sim \epsilon^{2}. \label{eqn:spincoeffsscaling}
\end{alignat}
These scalings match those found in \cite{Kunze:2004qd}, up to a change in convention for labeling the Newman-Penrose
tetrad. In order to compare with the aforementioned reference, relabel the spin coefficients according to the
tetrad relabeling $k \leftrightarrow l, m \leftrightarrow \overline{m}$. The directional derivatives in the
Newman-Penrose basis defined in \eqref{eqn:directionalderivatives} have the scaling
\begin{equation} \label{} 
D \sim \epsilon^{0}, \qquad \Delta \sim \epsilon^{-2}, \qquad \delta \sim \epsilon^{-1}, \qquad
\overline{\delta} \sim \epsilon^{-1}.
\end{equation}

So far in this appendix we have merely performed a coordinate transformation by introducing scaled adapted
coordinates $\left(\epsilon^{0} U, \epsilon^{2} V, \epsilon^{1} X^{i}\right)$. In order to have a well defined
Penrose limit, we will need to work with rescaled quantities. For instance, the $\epsilon \to 0$ limit of the
metric in scaled adapted coordinates is degenerate, as can be seen from \eqref{eqn:scaledmetric}. In this
appendix, we denote rescaled quantities with a circle above them. Thus, the rescaled metric is
\begin{equation} \label{} 
\mathring{ds}^{2} = \epsilon^{-2} ds^{2} = 2 dU dV + \epsilon^{2} D(x^{\mu}) dV^2 + 2 \epsilon^{1}
B_i(x^{\mu}) dV dX^i - C_{ij}(x^{\mu})dX^i dX^j \ .
\end{equation}
The Newman-Penrose tetrad corresponding to this rescaled metric is given by the rescaled version of the
Newman-Penrose tetrad $\left(k^{\mu}, n^{\mu}, E^{\mu}_{(a)}\right)$,
\begin{equation} \label{} 
\left( \mathring{k}^{\mu}, \mathring{n}^{\mu}, \mathring{E}^{\mu}_{(a)}\right) = \epsilon \left(
\epsilon^{-1} k^{\mu}, \epsilon n^{\mu}, E^{\mu}_{(a)}\right),
\end{equation}
where in addition to a $\sqrt{\epsilon^{2}}$ rescaling inherited from the (inverse) metric rescaling, we have applied a boost to $\left(k^{\mu}, n^{\mu}\right)$. This is done so that all elements of the rescaled tetrad
as well as the rescaled co-tetrad have a well-defined $\epsilon \to 0$ limit.
The leading order scaling of $\left(\mathring{k}^{\mu}, \mathring{n}^{\mu}, \mathring{E}^{\mu}_{(a)}\right)$, as
well as $\left(\mathring{k}_{\mu}, \mathring{n}_{\mu}, \mathring{E}_{\mu (a)}\right)$, is $\epsilon^{0}$
across the board, with subleading corrections containing only positive powers of  $\epsilon$. The same is
therefore true of the directional derivatives corresponding to this rescaled tetrad,
\begin{equation} \label{} 
\mathring{D} \sim \epsilon^{0}, \qquad \mathring{\Delta} \sim \epsilon^{0}, \qquad \mathring{\delta} \sim
\epsilon^{0}, \qquad \mathring{\overline{\delta}} \sim \epsilon^{0}.
\end{equation}

When building up solutions to field equations perturbatively in $\epsilon$ starting from the Penrose limit, there is a definite advantage
to using rescaled quantities that remain well-defined in the $\epsilon \to 0$ limit. The price is that the scaling of the original spin coefficients defined in \eqref{eqn:spincoeffssymbols} is somewhat obscured. With the rescaled
tetrad, we would naively expect all spin coefficients to have a leading order term that scales as $\epsilon^{0}$.
Nevertheless, it turns out that for the chosen Newman-Penrose tetrad, several of the spin coefficients vanish in the
Penrose limit and, in fact, the vanishing spin coefficients are precisely those that scale with a positive
power of $\epsilon$ in the original Newman-Penrose tetrad. This pattern continues at subleading orders, with
nonvanishing spin coefficients appearing only at the $\epsilon$ order where they did for the original
Newman-Penrose tetrad. Thus, the leading order scalings for the spin coefficients laid out in
\eqref{eqn:spincoeffsscaling} continues to hold for the rescaled tetrad.

\paragraph{Schwarzschild black hole in adapted coordinates} 

We provide here some further details on the adapted coordinates and coordinate transformations that are discussed in section \ref{sec:example:Schwarzschild} in the main text, for the example of Schwarzschild black hole. We do so to be self-contained but a pedagogic derivation of the adapted coordinates is already given in \cite{blau2011plane} and most of the results below more generally, in our conventions, are already presented in \cite{Chawla:2024mse}. Therefore, we omit derivations of the results.

First, we present the coordinate transformation between the Schwarzschild-Kerr-Schild metric 
\begin{equation}\label{eqn:app:SmetricKS}
ds^2 = d\hat{t}^2- dr^2-r^2 \left(d\theta^2 + \sin^2\theta d \phi^2\right) -\frac{2M}{r}(d\hat{t} + dr)^2  \, .
\end{equation}
and the adapted coordinates \eqref{eqn:Sadapted}
\begin{equation}\label{eqn:app:Sadapted}
\begin{aligned}
ds^2 =  
2 dU dV&- \left(r^2 \frac{\hat{E}^2}{L^2}-\left(1-\frac{2M}{r}\right)\right)d\hat{t}_0^2 - r^2 \sin^2(\theta) d \phi_0^2  \\ &-\frac{r^2}{L^2} dV^2+2\hat{E} \frac{r^2}{L^2} dV d\hat{t}_0 \, .\\
\end{aligned}
\end{equation}
The coordinate transformation between \eqref{eqn:app:SmetricKS} and \eqref{eqn:app:Sadapted} is given by \cite{Chawla:2024mse}
\begin{equation}\label{eqn:app:Schwarzschildadapted}
\begin{aligned}
U &= \int^{r(U)}_{r_{\rm ref}} \frac{d\rho}{R'(\rho) -\frac{2M}{\rho} \left(\hat{E}+R'(\rho)\right)}   \, , \\ 
\hat{t}(U,\hat{t}_0) &= \hat{t}_0 + \int^{r(U)}_{r_{\rm ref}} d\rho \, \frac{\hat{E}+\frac{2M}{\rho} \left(\hat{E}+ R'(\rho)\right)}{R'(\rho) -\frac{2M}{\rho} \left(\hat{E}+R'(\rho)\right)} \, , \\
\theta_0(V,\hat{t}_0) &= \frac{1}{L}\left(\hat{E}\hat{t}_0  - V\right) \, , \\
\theta(U,V, \hat{t}_0) &= \theta_0(V,\hat{t}_0)+ L\int^{r(U)}_{r_{\rm ref}} d\rho \, \frac{1}{\rho^2}\frac{1}{R'(\rho) -\frac{2M}{\rho} \left(\hat{E}+R'(\rho)\right)} \, ,\\
\phi(U,V,\phi_0,\hat{t}_0) &= \phi_0  \, ,
\end{aligned}
\end{equation}
where $r_{\rm ref}$ is an arbitrary constant while $r(U) = r(U;\hat{E},L^2)$ is a radial solution to the geodesic equations. Explicitly, $r(U)$ satisfies
\begin{equation}
\frac{dr}{dU} = \left(1-\frac{2M}{r(U)}\right)R'(r(U)) - \frac{2M \hat{E}}{r(U)} \, , 
\end{equation}
with initial condition $r(0) = r_{\rm ref}$ and where $R(r)$ satisfies
\begin{equation}\label{eqn:uschwarzschild}
\begin{aligned}
-L^2 &= -r^2 \hat{E}^2+r^2 \left(R'(r)\right)^2-2 M r \left(R'(r)+\hat{E}\right)^2 \, . \\
\end{aligned}
\end{equation}
In addition, $\hat{E}$ is the energy of the geodesic with respect to Kerr-Schild time and $L$ is the total angular momentum. The coordinates \eqref{eqn:app:Sadapted} are adapted to a congruence of geodesics which have no angular momentum along the $\phi$-direction; $L_{\phi} = 0$.

There are two branches of solution to \eqref{eqn:uschwarzschild}, 
\begin{equation}
R'(r) = \frac{2 \hat{E} M}{r-2M} \pm \frac{r}{r-2M} \sqrt{ \hat{E}^2-\frac{L^2}{r^2}(1-\frac{2M}{r})} \, , 
\end{equation}
corresponding respectively to an outgoing and an ingoing null geodesic congruence. Below, for definiteness, we consider the outgoing branch.

Using the coordinate transformation \eqref{eqn:app:Schwarzschildadapted} on flat space in Kerr-Schild coordinates
\begin{equation}\label{eqn:app:SmetricKSflat}
ds^2_{\flat} = d\hat{t}^2- dr^2-r^2 \left(d\theta^2 + \sin^2\theta d \phi^2\right)  \, ,
\end{equation}
and taking the Penrose limit we find \eqref{eqn:Sadaptedpenroseflat}
\begin{equation}\label{eqn:app:Sadaptedpenroseflat}
\begin{aligned}
ds^2_{\gamma,\flat} =  
2 dU dV&- \left(r^2 \frac{\hat{E}^2}{L^2}-1\right)d\hat{t}_0^2 - r^2 \sin^2 \theta d \phi_0^2 \\ &+ \frac{4 M \hat{E}}{r\left(1-\frac{2M}{r}\right)}\left(1+\sqrt{1-\frac{L^2}{\hat{E}^2 r^2}\left(1-\frac{2M}{r}\right)}\right)dU d\hat{t}_0 \\ &+ \frac{2M \hat{E}^2}{r\left(1-\frac{2M}{r}\right)}  \left(2-\frac{L^2}{\hat{E}^2 r^2}\left(1-\frac{2M}{r}\right)+2\sqrt{1-\frac{L^2}{\hat{E}^2 r^2}\left(1-\frac{2M}{r}\right)}\right) dU^2 \, .
\end{aligned}
\end{equation}

Finally, in the main text we have expressed \eqref{eqn:app:Sadaptedpenroseflat} in terms of the Brinkmann coordinates; we present the details here. The Penrose limit of \eqref{eqn:app:Sadapted}, which is given in Rosen coordinates by
\begin{equation}\label{eqn:SchwarzschildPenrose}
\begin{aligned}
ds^2_{\gamma} =  
2 dU dV&- \left(r^2 \frac{\hat{E}^2}{L^2}-\left(1-\frac{2M}{r}\right)\right)d\hat{t}_0^2 - r^2 \sin^2(\theta) d \phi_0^2   \, ,
\end{aligned}
\end{equation}
can be transformed to Brinkmann coordinates using \eqref{eqn:rosenBrinkmann} for the obvious choice of diagonal frame. We find 
\begin{equation}\label{eqn:app:rosenBrinkmann}
\begin{aligned}
U &= u  , \\
\hat{t}_0 &= \frac{L x_1}{r \sqrt{ \hat{E}^2-\frac{L^2}{r^2}(1-\frac{2M}{r})}}  \, ,\\
\phi_0 &= \frac{x_2}{r \sin \theta} \, , \\
V &= v - \frac{1}{4}\partial_u \log\left(r^2 \frac{\hat{E}^2}{L^2}-\left(1-\frac{2M}{r}\right)\right) x_1^2 - \frac{1}{4}\partial_u \log\left(r^2 \sin^2\theta\right)x_2^2 \, . 
\end{aligned}
\end{equation}
With this coordinate change, \eqref{eqn:SchwarzschildPenrose} becomes 
\begin{equation}\label{eqn:brinkmann}
ds^2_{\gamma} = 2 du dv + \frac{3 L^2 M}{r^5}\left(x_2^2-x_1^2\right)du^2 - dx_1^2-dx_2^2 \, .
\end{equation}
On the other hand, performing the same change of coordinates for the Penrose limit of the ``Kerr-Schild flat space'':
\begin{equation}
ds^2_{\flat} = ds^2 - \phi k_{\mu} k_{\nu}dx^{\mu}dx^{\nu} \, .
\end{equation}
In adapted coordinates \eqref{eqn:app:Sadaptedpenroseflat}, we find
\begin{equation}\label{eqn:app:Sadaptedpenroseflatbrinkmann}
\begin{aligned}
ds^2_{\gamma,\flat} &=  
2 du dv -dx^2_1 - dx_2^2\, \\
&+ 2 M x_1  \frac{2 L^2 (L^2 M - \hat{E}r^3) du dx_1}{r^{5/2} \left(2 L^2 M -L^2 r + \hat{E} r^3 \right)^{3/2}}  + \frac{2L^2 M}{2 L^2 M -L^2 r + \hat{E}^2 r^3} dx_1^2 \\
&+ \left( \frac{3 L^2 M x_2^2}{r^5} + \frac{2M \hat{E}^2}{r} X_0(u) + \frac{4 M L \hat{E} x_1}{r^3} X_1(u) + \frac{M L^2 x_1^2}{r^5} X_2(u) \right)du^2 \, ,
\end{aligned}
\end{equation}
which is \eqref{eqn:Sadaptedpenroseflatbrinkmann} in the main text with
\begin{equation}
\begin{aligned}
X_0(u)  &= \frac{L^2}{r \hat{E}^2(r-2M)}- \frac{2 r^2}{(r-2M)^2}+  \frac{2}{r (r-2M)^2}\sqrt{ 1-\frac{L^2}{\hat{E}^2 r^2}\left(1-\frac{2M}{r}\right)} \, ,\\
X_1(u)  &= \frac{\left(L^2 M-\hat{E}^2 r^3\right) \left(\sqrt{\hat{E}^2 r^3+2 L^2 M-L^2
		r}+\hat{E} r^{3/2}\right)}{\hat{E} \sqrt{r} (r-2 M) \left(\hat{E}^2 r^3+2 L^2
	M-L^2 r\right)} \, ,\\
X_2(u)  &= \frac{-r^2 \left(\hat{E}^2 r \left(\hat{E}^2 r^3+16 L^2 M-6 L^2 r\right)+3
	L^4\right)-10 L^4 M^2+12 L^4 M r}{\left(\hat{E}^2 r^3+2 L^2 M-L^2 r\right)^2} \, .\\
\end{aligned}
\end{equation}

\section{The Fackerell-Ipser equation}\label{app:FIequation}

In this appendix, we briefly discuss the Fackerell-Ipser equation and how it relates to \eqref{eqn:eomscalarcurved}
\begin{equation}\label{eqn:app:eomscalarcurved}
\left(\square  +  \sqrt{2 \Psi_{ABCD} \Psi^{ABCD}/3}\right) S = 0\, .
\end{equation}
First let us note that the invariant
\begin{equation}
I = \frac{1}{2}\Psi_{ABCD} \Psi^{ABCD} \, , 
\end{equation}
is often introduced in the context of determining the Petrov classification of a spacetime \cite{Stephani:2003tm}. In terms of the Weyl scalars in an arbitrary frame we have
\begin{equation}\label{eqn:IinWeyl}
\Psi_{ABCD} \Psi^{ABCD} = 2\Psi_0 \Psi_4 - 8\Psi_1 \Psi_3 + 6 \Psi_2^2 \, .
\end{equation}
Finally, in terms of the (bivector) eigenvalue problem (for say eigenvalue $\lambda$ and eigenbivector $X^{\alpha \beta}$) of the Weyl tensor
\begin{equation}
\frac{1}{2}C^{\mu \nu}{}_{\alpha \beta} X^{\alpha \beta} = \lambda X^{\mu \nu} \, ,
\end{equation}
whose eigenvalues are $\lambda_1$, $\lambda_2$, and $\lambda_3$, we have
\begin{equation}
\Psi_{ABCD} \Psi^{ABCD} = \lambda_1^2 + \lambda^2_2 + \lambda_3^2 \, .
\end{equation}
One can thus think of \eqref{eqn:app:eomscalarcurved} as a variation of the conformally coupled scalar but with a coupling to a type of matrix norm of the Weyl tensor instead of the Ricci scalar. 

The Fackerell-Ipser equation was introduced in (and named after) \cite{Fackerell:1972hg}, in the context of electromagnetic perturbations of a Kerr black hole. Specifically, it is the single partial differential equation that governs the zero GHP-weight component ($f_1$) of these perturbations with respect to a principal null frame. In that context, unlike the Teukolsky equations for the extremal GHP-weight components \cite{Teukolsky:1973ha}, the Fackerell-Ipser equation is not separable.

The derivation in \cite{Fackerell:1972hg}, starts from the general form of the Newman-Penrose Maxwell equations
\begin{align}
D f_{1} - \overline{\delta} f_{0} &= -\left(-\tau' + 2 \beta'\right) f_{0} - 2 \rho f_{1} + \kappa f_{2}, \label{eqn:fpertgen1}\\
D f_{2} - \overline{\delta} f_{1} &= -\sigma' f_{0} + 2 \tau' f_{1} - \left(\rho - 2 \varepsilon\right) f_{2}, \label{eqn:fpertgen2} \\
\delta f_{1} - \Delta f_{0} &= -\left(-\rho' + 2 \varepsilon' \right) f_{0} - 2 \tau f_{1} + \sigma f_{2}, \label{eqn:fpertgen3} \\
\delta f_{2} - \Delta f_{1} &= -\kappa' f_{0}  + 2 \rho' f_{1} - \left(\tau - 2 \beta\right) f_{2}, \label{eqn:fpertgen4}
\end{align}
which were also used to derive \eqref{eqn:feqnspert1}-\eqref{eqn:feqnspert4}. Then, using the further simplifications that occur for a Kerr spacetime using a principle null frame (which could be generalized to vacuum type D spacetimes more generally), \cite{Fackerell:1972hg} finds that, either by eliminating $f_{0}$ from the (Petrov D simplified) first two equations of \eqref{eqn:fpertgen1}-\eqref{eqn:fpertgen2} or by eliminating $f_{2}$ from the second two equations of \eqref{eqn:fpertgen3}-\eqref{eqn:fpertgen4}, we find
\begin{equation}\label{eqn:FIequation}
\left(\square  + 2 \Psi_2 \right) (\zeta f_1) = \left(\square  +  \sqrt{2\Psi_{ABCD} \Psi^{ABCD}/3} \right) (\zeta f_1)   = 0\, ,
\end{equation}
where $\Psi_2$ is evaluated in the principal null frame, using \eqref{eqn:IinWeyl} for the first equality in this special frame. In addition, we can define $\zeta$ through 
\begin{equation}
\Psi_2 = -\frac{M}{\zeta^3} \, .
\end{equation}
Now it can be straightforwardly observed that with the Weyl double copy relation for a type D spacetime in the principal null frame (see \eqref{eqn:scalarsdcrelations} with only $\Psi_2$ non-zero)
\begin{equation}
\Psi_2 = \frac{2 (f_1)^2}{S} \, , 
\end{equation}
holds for \cite{Luna:2018dpt}
\begin{equation}
\Psi_2 = -\frac{M}{\zeta^3} \, , \quad  f_1 =  -\frac{M}{\zeta^2}, \quad S = -\frac{2 M}{\zeta}\, .
\end{equation}
Moreover\footnote{Different choices of normalizations exist, in which case \eqref{eqn:typeDftoS} would hold only up to a proportionality constant. Of course, the Fackerell-Ipser equation is linear so it would still hold for $S$.}, 
\begin{equation}\label{eqn:typeDftoS}
2 \zeta f_1 = S \, ,
\end{equation}
satisfies \eqref{eqn:FIequation} on account of the fact that $f_I$ satisfy the Maxwell field equations, and thus $\zeta f_1$ satisfies the Fackerell-Ipser equation.

We have thus shown that the Fackerell-Ipser equation holds for the zeroth copy scalar of Petrov type D spacetimes in the Weyl double copy. By contrast, such a zeroth copy scalar generically does not satisfy the curved massless wave equation. For type N spacetimes, the Fackerell-Ipser equation reduces to a massless wave equation; in that case there is no distinction. In addition, we can formulate the Fackerell-Ipser equation for any spacetime and it is invariant with respect to frame rotations connected to the identity. For all these reasons, we have proposed in the main text that it is the natural field equation satisfied by the zeroth copy. One remaining subtlety that would be interesting to clarify is about the complex nature of the equation, which as formulated is not real, and, relatedly, the branch-choice of the square root.

\section{Field equations: solutions and effective sources}\label{app:leadingequations}

In Section \ref{eqn:curvedlocal}, we have discussed a particular class of solutions to the wave equations order-by-order on the leading order plane wave background. Specifically, we have made the ansatz \eqref{eqn:Sexpansion}-\eqref{eqn:fexpansion}, which is expected to arise from the Penrose expansion of a full solution. On the other hand, the plane wave equations of motion are simple enough to describe the full solutions which we will do here explicitly for the scalar equation, while indicating how a similar calculation would work for the Maxwell equations. We then give an example of the source terms at second subleading order. \\

First, in order to describe the scalar wave equation in full generality it is useful to note that the inverse of \eqref{eqn:adapted} is given by
\begin{equation}
g^{\mu \nu}\partial_{\mu}\partial_{\nu} = 2 \partial_U \partial_V - \left(D + B_i C^{ij} B_j\right)\partial^2_U+ 2 B^i \partial_U \partial_i- C^{ij}\partial_i \partial_j \, .
\end{equation}
Expanding the associated d'Alembertian order-by-order in the Penrose limit, we obtain at order $k$ an equation of the form \eqref{eqn:Seqnspertexpl}
\begin{equation}
\begin{aligned}
\left( \frac{1}{\sqrt{C_{(\gamma)}}}\partial_U \sqrt{C_{(\gamma)}} \partial_V + \partial_V\partial_U  - C_{(\gamma)}^{ij}  \partial_i  \partial_j \right)S^{(k)} = J_S^{(k)} \, .
\end{aligned}
\end{equation}
with (formal) solutions
\begin{equation}\label{eqn:adaptedscalarsolformal}
\begin{aligned}
S^{(i)} &= \int dk_V dk_i \, e^{i V k_V -i X^i k_i}\tilde{S}^{(i)}_{k_V, k_i}(U) \, , \quad J_S^{(i)} = \int dk_V dk_i \, e^{i V k_V -i X^i k_i}\tilde{J}^{(i)}_{k_V, k_i}(U) \\
\tilde{S}^{(i)}_{k_V, k_i}(U) &= \tilde{S}^{h \, (i)}_{k_V, k_i}(U) - \frac{i}{2 k_V} \int^U_{U_0} dU'' \left(\frac{C(U'')}{C(U)}\right)^{1/4} e^{\frac{i k_i k_j}{2 k_V } \int_{U''}^{U} dU' \, C^{ij}(U')} \tilde{J}_{k_V, k_i}^{(i)}(U'') \\
\tilde{S}^{h \, (i)}_{k_V, k_i}(U) &= c_{k_V, k_i} \frac{e^{\frac{i k_i k_j}{2 k_V} \int_{U_0}^U dU' \, C^{ij}(U')}}{C^{1/4}(U)} \, ,
\end{aligned}
\end{equation}
for (in principle) arbitrary $c_{k_V, k_i}$, fixed by, say, initial conditions for each $\tilde{S}^{(i)}_{k_V, k_i}(U)$. Naturally, the formal solution \eqref{eqn:adaptedscalarsolformal} could be ill-defined. Indeed, it is clear that $k_V \to 0$ will potentially be problematic and from the Penrose limit result, where $S^{(0)}(x^{\mu}) = S^{(0)}(U)$, this is in fact relevant. As a result, the full formal solution \eqref{eqn:adaptedscalarsolformal} is not particularly useful in the analysis in the main text. \\

For electro-magnetic perturbations, we instead wish to solve \eqref{eqn:feqnspert1}-\eqref{eqn:feqnspert4}
\begin{align}
\partial_U f^{(k)}_{1} + \frac{1}{\sqrt{2}}(E^i_{(1)}\partial_{X^1}+i E^i_{(2)} \partial_{X^i}) f^{(k)}_{0} + 2 \rho^{(0)} f^{(k)}_{1} &=  J^{(k)}_{f_1}[f^{0}_I, \ldots f^{k-1}_I],  \\
\partial_U f^{(k)}_{2}+ \frac{1}{\sqrt{2}}(E^i_{(1)}\partial_{X^1}+i E^i_{(2)} \partial_{X^i}) f^{(k)}_{1} +  \left(\rho^{(0)} - 2 \varepsilon^{(0)}\right) f^{(k)}_{2} &= J^{(i)}_{f_2}[f^{0}_I, \ldots f^{k-1}_I],  \\
\frac{1}{\sqrt{2}}(-E^i_{(1)}\partial_{X^1}+i E^i_{(2)} \partial_{X^i}) f^{(k)}_{1} - \partial_V f^{(k)}_{0}- \sigma^{(0)} f^{(k)}_{2}&=  J^{(k)}_{f_3}[f^{0}_I, \ldots f^{i-1}_I],  \\
\frac{1}{\sqrt{2}}(-E^i_{(1)}\partial_{X^1}+i E^i_{(2)} \partial_{X^i}) f^{(k)}_{2} - \partial_V f^{(k)}_{1} &=   J^{(k)}_{f_4}[f^{0}_I, \ldots f^{k-1}_I] . 
\end{align}
Here, we will not use the specific forms of the non-trivial (leading order) spin-coefficients $\rho^{(0)}$, $\epsilon^{(0)}$, and $\sigma^{(0)}$ but use that they are only functions of $U$. As a result, we will not write down solutions as explicitly as for the scalar case but it is still clear that after going to Fourier-space for $X^i$ and $V$, we are left with two  coupled first order linear ordinary differential equations in $\tilde{f}^{(k)}_{1,k_V,k_i}$ and $\tilde{f}^{(k)}_{2,k_V,k_i}$ together with two (algebraic) constraints.

\paragraph{Second order source terms}

We present explicitly the sources $T^{(1)}(U)$ as defined and discussed around \eqref{eqn:Meqns}
\begin{equation}\label{eqn:app:Meqns}
\begin{aligned}
M^{(1)}(U)\begin{pmatrix}
a^{(1)}_{S,0, 1,0}(U) \\ a^{(1)}_{S,0, 0,1}(U) 
\end{pmatrix} = T^{(1)}(U)  \, . \quad	
\end{aligned}
\end{equation}
As noted in the main text, the expressions are not particularly insightful, but we present them here as they are important to our conclusion that we do not generically expect the double copy to hold to second order in the Penrose expansion.

First, define the components of the vectors $T^{(1)}(U)$ 
\begin{equation}
T^{(1)}(U)  = \begin{pmatrix}
T_{1}^{(1)}(U)  \\ T_{2}^{(1)}(U) 
\end{pmatrix}  \, . 
\end{equation}
We find for $T^{(1)}(U)$
\begin{equation}
\begin{aligned}
T_{1}^{(1)}(U) &= \frac{a^{(0)}_{S,0,0,0}}{\sqrt{a^{(0)}_{\Psi_0,0,0,0} a^{(0)}_{S,0,0,0}}} \left( -\frac{da^{(1)}_{\Psi_1,0,0,0}}{dU}   -  a^{(1)}_{\Psi_1,0,0,0} \frac{d \log a^{(0)}_{S,0,0,0}}{2 dU}+a^{(1)}_{\Psi_1,0,0,0} \frac{d\log C}{2 dU} \right. \\ & \left. +\frac{i \bar{m}_{1} }{2 \sqrt{C}}  a^{(1)}_{\Psi_0,0,0,1}  -\frac{i \bar{m}_{2}}{2 \sqrt{C}}  a^{(1)}_{\Psi_0,0,1,0}+a^{(1)}_{\Psi_1,0,0,0}\frac{d \log a^{(0)}_{\Psi_0,0,0,0}}{2 dU}+\frac{i}{\sqrt{C}} (\partial_{1}\bar{m}_{2}-\partial_{2}\bar{m}_{1})a^{(0)}_{\Psi_0,0,0,0} \right)  \, ,
\end{aligned}
\end{equation}
and for $T_{2}^{(1)}(U)$
\begin{equation}
\begin{aligned}
T_{2}^{(1)}(U) &= \frac{i a^{(0)}_{S,0,0,0}}{\sqrt{C a^{(0)}_{\Psi_0,0,0,0} a^{(0)}_{S,0,0,0}}} \left\lbrack   \bar{m}_{1} a^{(2)}_{\Psi_1,0,0,1}-\bar{m}_{2} a^{(2)}_{\Psi_1,0,1,0} + a^{(2)}_{\Psi_2,0,0,0} \left(\bar{m}_{2}\partial_{U}m_{1} - \bar{m}_{1}\partial_{U}m_{2}\right) \right.  \\ &+ \frac{i a^{(1)}_{\Psi_1,0,0,0}}{\sqrt{C}}  \left(\bar{m}_{2} m_{1}-\bar{m}_{1} m_{2}\right)   \left(\bar{m}_{2} \partial_{U} B_1-\bar{m}_{1} \partial_{U} B_2\right) -2 i \sqrt{C} \frac{\partial_U (a^{(1)}_{\Psi_1,0,0,0})^2}{ a^{(0)}_{\Psi_0,0,0,0}}  \\ & +\frac{i \sqrt{C}}{2} \left(2  \partial_{U} a^{(2)}_{\Psi_2,0,0,0}{}+a^{(2)}_{\Psi_2,0,0,0} \partial_{U} \log a^{(0)}_{S,0,0,0}\right)  +  \frac{a^{(1)}_{\Psi_1,0,0,0}}{2 a^{(0)}_{\Psi_0,0,0,0}}   \left(\bar{m}_{1} a^{(1)}_{\Psi_0,0,0,1}-\bar{m}_{2} a^{(1)}_{\Psi_0,0,1,0}\right) \\ &+4 \frac{a^{(1)}_{\Psi_1,0,0,0}}{2 a^{(0)}_{\Psi_0,0,0,0}} (\partial_{U}m_{2} \bar{m}_{1}-\partial_{U}m_{1} \bar{m}_{2}) a^{(1)}_{\Psi_1,0,0,0}  - \frac{i \sqrt{C}a^{(2)}_{\Psi_2,0,0,0}}{2 a^{(0)}_{\Psi_0,0,0,0}}   \partial_U a^{(0)}_{\Psi_0,0,0,0} \\ &-2 i \sqrt{C} (a^{(1)}_{\Psi_1,0,0,0})^2 \partial_U \log a^{(0)}_{S,0,0,0} +\frac{3 i \sqrt{C} (a^{(1)}_{\Psi_1,0,0,0})^2}{a^{(0)}_{\Psi_0,0,0,0}} \partial_U \log a^{(0)}_{\Psi_0,0,0,0} \\ &  +\frac{a^{(0)}_{\Psi_0,0,0,0}}{C} \left(\bar{m}_{1} m_{2}-\bar{m}_{2} m_{1}\right)  \left(\bar{m}_{1}^2 \partial_{X^2} B_2-\bar{m}_{2} \bar{m}_{1} \left(\partial_{X^2} B_1+\partial_{X^1} B_2\right) +\bar{m}^2_{2} \partial_{X^1} B_1 \right) \\ & \left. + \frac{a^{(0)}_{\Psi_0,0,0,0}}{C} \left(\bar{m}_{1} m_{2}-\bar{m}_{2} m_{1}\right) (\bar{m}_{1} \partial_{V}\bar{m}_{2} - \bar{m}_{2}\partial_{V}\bar{m}_{1} ) (\bar{m}_{1} m_{2}-\bar{m}_{2} m_{1}) \right\rbrack \\ 
&+ \frac{\sqrt{a^{(0)}_{\Psi_0,0,0,0} a^{(0)}_{S,0,0,0}}}{C^{3/2}}   \left(t_{2,D}^{(1)} D(U) + \sum^2_{i=1}\left(t_{2,B_i}^{(1)} B_i+ \sum^2_{j=1} t_{2,B_i B_j }^{(1)} B_i B_j \right)\right)\, , 
\end{aligned}
\end{equation}
where
\begin{equation}
\begin{aligned}
t_{2,D}^{(1)} &= -\frac{i}{2} (\bar{m}_{1} \partial_{U}\bar{m}_{2} - \bar{m}_{2} \partial_{U}\bar{m}_{1} ) (\bar{m}_{2} m_{1}-\bar{m}_{1} m_{2})^2  \, , \\
t_{2,B_1}^{(1)} &=  \frac{\sqrt{C} a^{(1)}_{\Psi_1,0,0,0}}{a^{(0)}_{\Psi_0,0,0,0}} \left( \frac{m_{1}}{2} \partial_{U}\bar{m}^2_{2} -3 \bar{m}_{2} m_{2} \partial_{U}\bar{m}_{1}  +\bar{m}_{2}^2 \partial_{U}m_{1} + \bar{m}_{1} \left(2 m_{2} \partial_{U}\bar{m}_{2}- \bar{m}_{2}\partial_{U}m_{2}\right) \right) \\ &-i  \left( \bar{m}_{2}^2 m_{1} (\partial_{X^2}\bar{m}_{1}-\partial_{X^1}\bar{m}_{2})+ \bar{m}_{2}^2 \bar{m}_{1} (\partial_{X^1}m_{2}+\partial_{X^2}m_{1}) \right) \\ &-i \left(\bar{m}_{2}^2 \left( m_{2}\partial_{X^1}\bar{m}_{1} -\bar{m}_{2} \partial_{X^1}m_{1}\right)+ \bar{m}_{1}^2\left( m_{2}\partial_{X^2}\bar{m}_{2} - \bar{m}_{2}\partial_{X^2}m_{2} \right)-2 m_{2} \bar{m}_{2} \bar{m}_{1} \partial_{X^2}\bar{m}_{1} \right) \, , \\
t_{2,B_2}^{(1)} &= \frac{\sqrt{C} a^{(1)}_{\Psi_1,0,0,0}}{a^{(0)}_{\Psi_0,0,0,0}} \left(\frac{m_{2}}{2}\partial_{U}\bar{m}^2_{1} -3 \bar{m}_{1} m_{1} \partial_{U}\bar{m}_{2}+ \bar{m}_{1}^2\partial_{U}m_{2} + \bar{m}_{2}\left( 2 m_{1}  \partial_{U}\bar{m}_{1} -\bar{m}_{1}\partial_{U}m_{1}   \right) \right) \\ &- i \left( \bar{m}_{1}^2 m_{2} (\partial_{X^2}\bar{m}_{1}-\partial_{X^1}\bar{m}_{2})- \bar{m}_{1}^2 \bar{m}_{2} \left(\partial_{X^1}m_{2}+\partial_{X^2}m_{1}\right)\right) \\ &- i \left(\bar{m}_{1}^2 \left(\bar{m}_{1} \partial_{X^2}m_{2}- m_{1} \partial_{X^2}\bar{m}_{2} \right) + \bar{m}_{2}^2 \left(\bar{m}_{1} \partial_{X^1}m_{1}-  m_{1}\partial_{X^1}\bar{m}_{1} \right) + 2 \bar{m}_{2} \bar{m}_{1} m_{1}\partial_{X^1}\bar{m}_{2}\right)  \, , 
\end{aligned}
\end{equation}
\begin{equation}
\begin{aligned}
t_{2,B_1 B_1}^{(1)} &=  i \bar{m}_{2} m_{2} (\partial_{U}\bar{m}_{2} \bar{m}_{1}-\partial_{U}\bar{m}_{1} \bar{m}_{2})  \, , \\
t_{2,B_1 B_2}^{(1)} = t_{2,B_2 B_1}^{(1)}  &= - \frac{i}{2} (\partial_{U}\bar{m}_{2} \bar{m}_{1}-\partial_{U}\bar{m}_{1} \bar{m}_{2}) (\bar{m}_{2} m_{1}+\bar{m}_{1} m_{2})  \, , \\
t_{2,B_2 B_2}^{(1)} &= i \bar{m}_{1}^2 \partial_{U}\bar{m}_{2}  m_{1} -2 i \bar{m}_{1} \bar{m}_{2} m_{1} \partial_{U}\bar{m}_{1} \, .
\end{aligned}
\end{equation}
Here, all quantities are implicitly evaluated at $X^i = V = 0$. For instance, $B_i = B_i(U,0,0,0)$ etc. Note that, despite the notation, $T_{2}^{(1)}(U)$ is a second order quantity. There is no obstruction to finding a first order solution to the equation involving $T_{1}^{(1)}(U)$ by itself. On the other hand, such solutions do not generically extend to higher orders. In that sense, this could be viewed as a ``linearization instability''.

\end{appendices}

\bibliographystyle{jhep}
\bibliography{bibfile} 

\providecommand{\href}[2]{#2}\begingroup\raggedright\begin{thebibliography}{100}

\bibitem{Bern:2008qj}
Z.~Bern, J.J.M.~Carrasco and H.~Johansson, \emph{{New Relations for Gauge-Theory Amplitudes}}, \href{https://doi.org/10.1103/PhysRevD.78.085011}{\emph{Phys. Rev. D} {\bfseries 78} (2008) 085011} [\href{https://arxiv.org/abs/0805.3993}{{\ttfamily 0805.3993}}].

\bibitem{Bern:2019nnu}
Z.~Bern, C.~Cheung, R.~Roiban, C.-H.~Shen, M.P.~Solon and M.~Zeng, \emph{{Scattering Amplitudes and the Conservative Hamiltonian for Binary Systems at Third Post-Minkowskian Order}}, \href{https://doi.org/10.1103/PhysRevLett.122.201603}{\emph{Phys. Rev. Lett.} {\bfseries 122} (2019) 201603} [\href{https://arxiv.org/abs/1901.04424}{{\ttfamily 1901.04424}}].

\bibitem{Bern:2019crd}
Z.~Bern, C.~Cheung, R.~Roiban, C.-H.~Shen, M.P.~Solon and M.~Zeng, \emph{{Black Hole Binary Dynamics from the Double Copy and Effective Theory}}, \href{https://doi.org/10.1007/JHEP10(2019)206}{\emph{JHEP} {\bfseries 10} (2019) 206} [\href{https://arxiv.org/abs/1908.01493}{{\ttfamily 1908.01493}}].

\bibitem{Bern:2020uwk}
Z.~Bern, J.~Parra-Martinez, R.~Roiban, E.~Sawyer and C.-H.~Shen, \emph{{Leading Nonlinear Tidal Effects and Scattering Amplitudes}}, \href{https://doi.org/10.1007/JHEP05(2021)188}{\emph{JHEP} {\bfseries 05} (2021) 188} [\href{https://arxiv.org/abs/2010.08559}{{\ttfamily 2010.08559}}].

\bibitem{Bern:2021yeh}
Z.~Bern, J.~Parra-Martinez, R.~Roiban, M.S.~Ruf, C.-H.~Shen, M.P.~Solon et~al., \emph{{Scattering Amplitudes, the Tail Effect, and Conservative Binary Dynamics at O(G4)}}, \href{https://doi.org/10.1103/PhysRevLett.128.161103}{\emph{Phys. Rev. Lett.} {\bfseries 128} (2022) 161103} [\href{https://arxiv.org/abs/2112.10750}{{\ttfamily 2112.10750}}].

\bibitem{Bern:2022jvn}
Z.~Bern, J.~Parra-Martinez, R.~Roiban, M.S.~Ruf, C.-H.~Shen, M.P.~Solon et~al., \emph{{Scattering amplitudes and conservative dynamics at the fourth post-Minkowskian order}}, \href{https://doi.org/10.22323/1.416.0051}{\emph{PoS} {\bfseries LL2022} (2022) 051}.

\bibitem{Bern:2019prr}
Z.~Bern, J.J.~Carrasco, M.~Chiodaroli, H.~Johansson and R.~Roiban, \emph{{The duality between color and kinematics and its applications}}, \href{https://doi.org/10.1088/1751-8121/ad5fd0}{\emph{J. Phys. A} {\bfseries 57} (2024) 333002} [\href{https://arxiv.org/abs/1909.01358}{{\ttfamily 1909.01358}}].

\bibitem{Kosower:2022yvp}
D.A.~Kosower, R.~Monteiro and D.~O'Connell, \emph{{The SAGEX review on scattering amplitudes Chapter 14: Classical gravity from scattering amplitudes}}, \href{https://doi.org/10.1088/1751-8121/ac8846}{\emph{J. Phys. A} {\bfseries 55} (2022) 443015} [\href{https://arxiv.org/abs/2203.13025}{{\ttfamily 2203.13025}}].

\bibitem{White:2024pve}
C.D.~White, \emph{{The Classical Double Copy}}, World Scientific (5, 2024), \href{https://doi.org/10.1142/q0457}{10.1142/q0457}.

\bibitem{Monteiro:2014cda}
R.~Monteiro, D.~O'Connell and C.D.~White, \emph{{Black holes and the double copy}}, \href{https://doi.org/10.1007/JHEP12(2014)056}{\emph{JHEP} {\bfseries 12} (2014) 056} [\href{https://arxiv.org/abs/1410.0239}{{\ttfamily 1410.0239}}].

\bibitem{Luna:2018dpt}
A.~Luna, R.~Monteiro, I.~Nicholson and D.~O'Connell, \emph{{Type D Spacetimes and the Weyl Double Copy}}, \href{https://doi.org/10.1088/1361-6382/ab03e6}{\emph{Class. Quant. Grav.} {\bfseries 36} (2019) 065003} [\href{https://arxiv.org/abs/1810.08183}{{\ttfamily 1810.08183}}].

\bibitem{Luna:2015paa}
A.~Luna, R.~Monteiro, D.~O'Connell and C.D.~White, \emph{{The classical double copy for Taub\textendash{}NUT spacetime}}, \href{https://doi.org/10.1016/j.physletb.2015.09.021}{\emph{Phys. Lett. B} {\bfseries 750} (2015) 272} [\href{https://arxiv.org/abs/1507.01869}{{\ttfamily 1507.01869}}].

\bibitem{Luna:2017dtq}
A.~Luna, I.~Nicholson, D.~O'Connell and C.D.~White, \emph{{Inelastic Black Hole Scattering from Charged Scalar Amplitudes}}, \href{https://doi.org/10.1007/JHEP03(2018)044}{\emph{JHEP} {\bfseries 03} (2018) 044} [\href{https://arxiv.org/abs/1711.03901}{{\ttfamily 1711.03901}}].

\bibitem{Ridgway:2015fdl}
A.K.~Ridgway and M.B.~Wise, \emph{{Static Spherically Symmetric Kerr-Schild Metrics and Implications for the Classical Double Copy}}, \href{https://doi.org/10.1103/PhysRevD.94.044023}{\emph{Phys. Rev. D} {\bfseries 94} (2016) 044023} [\href{https://arxiv.org/abs/1512.02243}{{\ttfamily 1512.02243}}].

\bibitem{Adamo:2017nia}
T.~Adamo, E.~Casali, L.~Mason and S.~Nekovar, \emph{{Scattering on plane waves and the double copy}}, \href{https://doi.org/10.1088/1361-6382/aa9961}{\emph{Class. Quant. Grav.} {\bfseries 35} (2018) 015004} [\href{https://arxiv.org/abs/1706.08925}{{\ttfamily 1706.08925}}].

\bibitem{Bahjat-Abbas:2017htu}
N.~Bahjat-Abbas, A.~Luna and C.D.~White, \emph{{The Kerr-Schild double copy in curved spacetime}}, \href{https://doi.org/10.1007/JHEP12(2017)004}{\emph{JHEP} {\bfseries 12} (2017) 004} [\href{https://arxiv.org/abs/1710.01953}{{\ttfamily 1710.01953}}].

\bibitem{CarrilloGonzalez:2018ejf}
M.~Carrillo~Gonz\'alez, R.~Penco and M.~Trodden, \emph{{Radiation of scalar modes and the classical double copy}}, \href{https://doi.org/10.1007/JHEP11(2018)065}{\emph{JHEP} {\bfseries 11} (2018) 065} [\href{https://arxiv.org/abs/1809.04611}{{\ttfamily 1809.04611}}].

\bibitem{Ilderton:2018lsf}
A.~Ilderton, \emph{{Screw-symmetric gravitational waves: a double copy of the vortex}}, \href{https://doi.org/10.1016/j.physletb.2018.04.069}{\emph{Phys. Lett. B} {\bfseries 782} (2018) 22} [\href{https://arxiv.org/abs/1804.07290}{{\ttfamily 1804.07290}}].

\bibitem{Gurses:2018ckx}
M.~Gurses and B.~Tekin, \emph{{Classical Double Copy: Kerr-Schild-Kundt metrics from Yang-Mills Theory}}, \href{https://doi.org/10.1103/PhysRevD.98.126017}{\emph{Phys. Rev. D} {\bfseries 98} (2018) 126017} [\href{https://arxiv.org/abs/1810.03411}{{\ttfamily 1810.03411}}].

\bibitem{Carrillo-Gonzalez:2017iyj}
M.~Carrillo-Gonz\'alez, R.~Penco and M.~Trodden, \emph{{The classical double copy in maximally symmetric spacetimes}}, \href{https://doi.org/10.1007/JHEP04(2018)028}{\emph{JHEP} {\bfseries 04} (2018) 028} [\href{https://arxiv.org/abs/1711.01296}{{\ttfamily 1711.01296}}].

\bibitem{Lee:2018gxc}
K.~Lee, \emph{{Kerr-Schild Double Field Theory and Classical Double Copy}}, \href{https://doi.org/10.1007/JHEP10(2018)027}{\emph{JHEP} {\bfseries 10} (2018) 027} [\href{https://arxiv.org/abs/1807.08443}{{\ttfamily 1807.08443}}].

\bibitem{Lescano:2020nve}
E.~Lescano and J.A.~Rodr\'\i{}guez, \emph{{$ \mathcal{N} $ = 1 supersymmetric Double Field Theory and the generalized Kerr-Schild ansatz}}, \href{https://doi.org/10.1007/JHEP10(2020)148}{\emph{JHEP} {\bfseries 10} (2020) 148} [\href{https://arxiv.org/abs/2002.07751}{{\ttfamily 2002.07751}}].

\bibitem{Lescano:2021ooe}
E.~Lescano and J.A.~Rodr\'\i{}guez, \emph{{Higher-derivative heterotic Double Field Theory and classical double copy}}, \href{https://doi.org/10.1007/JHEP07(2021)072}{\emph{JHEP} {\bfseries 07} (2021) 072} [\href{https://arxiv.org/abs/2101.03376}{{\ttfamily 2101.03376}}].

\bibitem{Bah:2019sda}
I.~Bah, R.~Dempsey and P.~Weck, \emph{{Kerr-Schild Double Copy and Complex Worldlines}}, \href{https://doi.org/10.1007/JHEP02(2020)180}{\emph{JHEP} {\bfseries 02} (2020) 180} [\href{https://arxiv.org/abs/1910.04197}{{\ttfamily 1910.04197}}].

\bibitem{Goldberger:2019xef}
W.D.~Goldberger and J.~Li, \emph{{Strings, extended objects, and the classical double copy}}, \href{https://doi.org/10.1007/JHEP02(2020)092}{\emph{JHEP} {\bfseries 02} (2020) 092} [\href{https://arxiv.org/abs/1912.01650}{{\ttfamily 1912.01650}}].

\bibitem{Kim:2019jwm}
K.~Kim, K.~Lee, R.~Monteiro, I.~Nicholson and D.~Peinador~Veiga, \emph{{The Classical Double Copy of a Point Charge}}, \href{https://doi.org/10.1007/JHEP02(2020)046}{\emph{JHEP} {\bfseries 02} (2020) 046} [\href{https://arxiv.org/abs/1912.02177}{{\ttfamily 1912.02177}}].

\bibitem{Luna:2020adi}
A.~Luna, S.~Nagy and C.~White, \emph{{The convolutional double copy: a case study with a point}}, \href{https://doi.org/10.1007/JHEP09(2020)062}{\emph{JHEP} {\bfseries 09} (2020) 062} [\href{https://arxiv.org/abs/2004.11254}{{\ttfamily 2004.11254}}].

\bibitem{Easson:2020esh}
D.A.~Easson, C.~Keeler and T.~Manton, \emph{{Classical double copy of nonsingular black holes}}, \href{https://doi.org/10.1103/PhysRevD.102.086015}{\emph{Phys. Rev. D} {\bfseries 102} (2020) 086015} [\href{https://arxiv.org/abs/2007.16186}{{\ttfamily 2007.16186}}].

\bibitem{Berman:2018hwd}
D.S.~Berman, E.~Chac\'on, A.~Luna and C.D.~White, \emph{{The self-dual classical double copy, and the Eguchi-Hanson instanton}}, \href{https://doi.org/10.1007/JHEP01(2019)107}{\emph{JHEP} {\bfseries 01} (2019) 107} [\href{https://arxiv.org/abs/1809.04063}{{\ttfamily 1809.04063}}].

\bibitem{Alkac:2021seh}
G.~Alkac, M.K.~Gumus and M.A.~Olpak, \emph{{Kerr-Schild double copy of the Coulomb solution in three dimensions}}, \href{https://doi.org/10.1103/PhysRevD.104.044034}{\emph{Phys. Rev. D} {\bfseries 104} (2021) 044034} [\href{https://arxiv.org/abs/2105.11550}{{\ttfamily 2105.11550}}].

\bibitem{Mkrtchyan:2022ulc}
K.~Mkrtchyan and M.~Svazas, \emph{{Solutions in Nonlinear Electrodynamics and their double copy regular black holes}}, \href{https://doi.org/10.1007/JHEP09(2022)012}{\emph{JHEP} {\bfseries 09} (2022) 012} [\href{https://arxiv.org/abs/2205.14187}{{\ttfamily 2205.14187}}].

\bibitem{Adamo:2022rob}
T.~Adamo, A.~Cristofoli and P.~Tourkine, \emph{{The ultrarelativistic limit of Kerr}}, \href{https://doi.org/10.1007/JHEP02(2023)107}{\emph{JHEP} {\bfseries 02} (2023) 107} [\href{https://arxiv.org/abs/2209.05730}{{\ttfamily 2209.05730}}].

\bibitem{Alfonsi:2020lub}
L.~Alfonsi, C.D.~White and S.~Wikeley, \emph{{Topology and Wilson lines: global aspects of the double copy}}, \href{https://doi.org/10.1007/JHEP07(2020)091}{\emph{JHEP} {\bfseries 07} (2020) 091} [\href{https://arxiv.org/abs/2004.07181}{{\ttfamily 2004.07181}}].

\bibitem{Adamo:2020qru}
T.~Adamo and A.~Ilderton, \emph{{Classical and quantum double copy of back-reaction}}, \href{https://doi.org/10.1007/JHEP09(2020)200}{\emph{JHEP} {\bfseries 09} (2020) 200} [\href{https://arxiv.org/abs/2005.05807}{{\ttfamily 2005.05807}}].

\bibitem{Gonzo:2021drq}
R.~Gonzo and C.~Shi, \emph{{Geodesics from classical double copy}}, \href{https://doi.org/10.1103/PhysRevD.104.105012}{\emph{Phys. Rev. D} {\bfseries 104} (2021) 105012} [\href{https://arxiv.org/abs/2109.01072}{{\ttfamily 2109.01072}}].

\bibitem{Keeler:2020rcv}
C.~Keeler, T.~Manton and N.~Monga, \emph{{From Navier-Stokes to Maxwell via Einstein}}, \href{https://doi.org/10.1007/JHEP08(2020)147}{\emph{JHEP} {\bfseries 08} (2020) 147} [\href{https://arxiv.org/abs/2005.04242}{{\ttfamily 2005.04242}}].

\bibitem{Easson:2021asd}
D.A.~Easson, T.~Manton and A.~Svesko, \emph{{Sources in the Weyl Double Copy}}, \href{https://doi.org/10.1103/PhysRevLett.127.271101}{\emph{Phys. Rev. Lett.} {\bfseries 127} (2021) 271101} [\href{https://arxiv.org/abs/2110.02293}{{\ttfamily 2110.02293}}].

\bibitem{Easson:2022zoh}
D.A.~Easson, T.~Manton and A.~Svesko, \emph{{Einstein-Maxwell theory and the Weyl double copy}}, \href{https://doi.org/10.1103/PhysRevD.107.044063}{\emph{Phys. Rev. D} {\bfseries 107} (2023) 044063} [\href{https://arxiv.org/abs/2210.16339}{{\ttfamily 2210.16339}}].

\bibitem{Godazgar:2020zbv}
H.~Godazgar, M.~Godazgar, R.~Monteiro, D.~Peinador~Veiga and C.N.~Pope, \emph{{Weyl Double Copy for Gravitational Waves}}, \href{https://doi.org/10.1103/PhysRevLett.126.101103}{\emph{Phys. Rev. Lett.} {\bfseries 126} (2021) 101103} [\href{https://arxiv.org/abs/2010.02925}{{\ttfamily 2010.02925}}].

\bibitem{White:2020sfn}
C.D.~White, \emph{{Twistorial Foundation for the Classical Double Copy}}, \href{https://doi.org/10.1103/PhysRevLett.126.061602}{\emph{Phys. Rev. Lett.} {\bfseries 126} (2021) 061602} [\href{https://arxiv.org/abs/2012.02479}{{\ttfamily 2012.02479}}].

\bibitem{Chacon:2021wbr}
E.~Chac\'on, S.~Nagy and C.D.~White, \emph{{The Weyl double copy from twistor space}}, \href{https://doi.org/10.1007/JHEP05(2021)239}{\emph{JHEP} {\bfseries 05} (2021) 2239} [\href{https://arxiv.org/abs/2103.16441}{{\ttfamily 2103.16441}}].

\bibitem{Chacon:2021lox}
E.~Chac\'on, S.~Nagy and C.D.~White, \emph{{Alternative formulations of the twistor double copy}}, \href{https://doi.org/10.1007/JHEP03(2022)180}{\emph{JHEP} {\bfseries 03} (2022) 180} [\href{https://arxiv.org/abs/2112.06764}{{\ttfamily 2112.06764}}].

\bibitem{Han:2022mze}
S.~Han, \emph{{The Weyl double copy in vacuum spacetimes with a cosmological constant}}, \href{https://doi.org/10.1007/JHEP09(2022)238}{\emph{JHEP} {\bfseries 09} (2022) 238} [\href{https://arxiv.org/abs/2205.08654}{{\ttfamily 2205.08654}}].

\bibitem{Alawadhi:2020jrv}
R.~Alawadhi, D.S.~Berman and B.~Spence, \emph{{Weyl doubling}}, \href{https://doi.org/10.1007/JHEP09(2020)127}{\emph{JHEP} {\bfseries 09} (2020) 127} [\href{https://arxiv.org/abs/2007.03264}{{\ttfamily 2007.03264}}].

\bibitem{Godazgar:2021iae}
H.~Godazgar, M.~Godazgar, R.~Monteiro, D.~Peinador~Veiga and C.N.~Pope, \emph{{Asymptotic Weyl double copy}}, \href{https://doi.org/10.1007/JHEP11(2021)126}{\emph{JHEP} {\bfseries 11} (2021) 126} [\href{https://arxiv.org/abs/2109.07866}{{\ttfamily 2109.07866}}].

\bibitem{Armstrong-Williams:2023ssz}
K.~Armstrong-Williams and C.D.~White, \emph{{A spinorial double copy for $ \mathcal{N} $ = 0 supergravity}}, \href{https://doi.org/10.1007/JHEP05(2023)047}{\emph{JHEP} {\bfseries 05} (2023) 047} [\href{https://arxiv.org/abs/2303.04631}{{\ttfamily 2303.04631}}].

\bibitem{Armstrong-Williams:2024bog}
K.~Armstrong-Williams, N.~Moynihan and C.D.~White, \emph{{Deriving Weyl double copies with sources}}, \href{https://doi.org/10.1007/JHEP03(2025)121}{\emph{JHEP} {\bfseries 03} (2025) 121} [\href{https://arxiv.org/abs/2407.18107}{{\ttfamily 2407.18107}}].

\bibitem{Adamo:2023fbj}
T.~Adamo, G.~Bogna, L.~Mason and A.~Sharma, \emph{{Scattering on self-dual Taub-NUT}}, \href{https://doi.org/10.1088/1361-6382/ad12ee}{\emph{Class. Quant. Grav.} {\bfseries 41} (2024) 015030} [\href{https://arxiv.org/abs/2309.03834}{{\ttfamily 2309.03834}}].

\bibitem{Sivaramakrishnan:2021srm}
A.~Sivaramakrishnan, \emph{{Towards color-kinematics duality in generic spacetimes}}, \href{https://doi.org/10.1007/JHEP04(2022)036}{\emph{JHEP} {\bfseries 04} (2022) 036} [\href{https://arxiv.org/abs/2110.15356}{{\ttfamily 2110.15356}}].

\bibitem{Herderschee:2022ntr}
A.~Herderschee, R.~Roiban and F.~Teng, \emph{{On the differential representation and color-kinematics duality of AdS boundary correlators}}, \href{https://doi.org/10.1007/JHEP05(2022)026}{\emph{JHEP} {\bfseries 05} (2022) 026} [\href{https://arxiv.org/abs/2201.05067}{{\ttfamily 2201.05067}}].

\bibitem{Cheung:2022pdk}
C.~Cheung, J.~Parra-Martinez and A.~Sivaramakrishnan, \emph{{On-shell correlators and color-kinematics duality in curved symmetric spacetimes}}, \href{https://doi.org/10.1007/JHEP05(2022)027}{\emph{JHEP} {\bfseries 05} (2022) 027} [\href{https://arxiv.org/abs/2201.05147}{{\ttfamily 2201.05147}}].

\bibitem{Ilderton:2024oly}
A.~Ilderton and W.~Lindved, \emph{{Toward double copy on arbitrary backgrounds}}, \href{https://doi.org/10.1007/JHEP11(2024)100}{\emph{JHEP} {\bfseries 11} (2024) 100} [\href{https://arxiv.org/abs/2405.10016}{{\ttfamily 2405.10016}}].

\bibitem{Cheung:2022mix}
C.~Cheung, J.~Mangan, J.~Parra-Martinez and N.~Shah, \emph{{Non-perturbative Double Copy in Flatland}}, \href{https://doi.org/10.1103/PhysRevLett.129.221602}{\emph{Phys. Rev. Lett.} {\bfseries 129} (2022) 221602} [\href{https://arxiv.org/abs/2204.07130}{{\ttfamily 2204.07130}}].

\bibitem{Prabhu:2020avf}
S.G.~Prabhu, \emph{{The classical double copy in curved spacetimes: perturbative Yang-Mills from the bi-adjoint scalar}}, \href{https://doi.org/10.1007/JHEP05(2024)117}{\emph{JHEP} {\bfseries 05} (2024) 117} [\href{https://arxiv.org/abs/2011.06588}{{\ttfamily 2011.06588}}].

\bibitem{Alkac:2021bav}
G.~Alkac, M.K.~Gumus and M.~Tek, \emph{{The Kerr-Schild Double Copy in Lifshitz Spacetime}}, \href{https://doi.org/10.1007/JHEP05(2021)214}{\emph{JHEP} {\bfseries 05} (2021) 214} [\href{https://arxiv.org/abs/2103.06986}{{\ttfamily 2103.06986}}].

\bibitem{Han:2022ubu}
S.~Han, \emph{{Weyl double copy and massless free-fields in curved spacetimes}}, \href{https://doi.org/10.1088/1361-6382/ac96c2}{\emph{Class. Quant. Grav.} {\bfseries 39} (2022) 225009} [\href{https://arxiv.org/abs/2204.01907}{{\ttfamily 2204.01907}}].

\bibitem{Didenko:2022qxq}
V.E.~Didenko and N.K.~Dosmanbetov, \emph{{Classical Double Copy and Higher-Spin Fields}}, \href{https://doi.org/10.1103/PhysRevLett.130.071603}{\emph{Phys. Rev. Lett.} {\bfseries 130} (2023) 071603} [\href{https://arxiv.org/abs/2210.04704}{{\ttfamily 2210.04704}}].

\bibitem{He:2023iew}
J.-L.~He and J.-H.~Huang, \emph{{Cosmological horizons from classical double copy}}, \href{https://doi.org/10.1016/j.physletb.2024.138579}{\emph{Phys. Lett. B} {\bfseries 851} (2024) 138579} [\href{https://arxiv.org/abs/2312.00972}{{\ttfamily 2312.00972}}].

\bibitem{Kent:2024mow}
B.~Kent, T.~Manton and S.~Shashi, \emph{{Background ambiguity and the G\"odel double copy}}, \href{https://doi.org/10.1007/JHEP03(2025)033}{\emph{JHEP} {\bfseries 03} (2025) 033} [\href{https://arxiv.org/abs/2411.04207}{{\ttfamily 2411.04207}}].

\bibitem{Zhao:2024wtn}
W.~Zhao, P.-J.~Mao and J.-B.~Wu, \emph{{Weyl double copy in type D spacetime in four and five dimensions}}, \href{https://doi.org/10.1103/PhysRevD.111.066005}{\emph{Phys. Rev. D} {\bfseries 111} (2025) 066005} [\href{https://arxiv.org/abs/2411.04774}{{\ttfamily 2411.04774}}].

\bibitem{Anastasiou:2014qba}
A.~Anastasiou, L.~Borsten, M.J.~Duff, L.J.~Hughes and S.~Nagy, \emph{{Yang-Mills origin of gravitational symmetries}}, \href{https://doi.org/10.1103/PhysRevLett.113.231606}{\emph{Phys. Rev. Lett.} {\bfseries 113} (2014) 231606} [\href{https://arxiv.org/abs/1408.4434}{{\ttfamily 1408.4434}}].

\bibitem{LopesCardoso:2018xes}
G.~Lopes~Cardoso, G.~Inverso, S.~Nagy and S.~Nampuri, \emph{{Comments on the double copy construction for gravitational theories}}, \href{https://doi.org/10.22323/1.318.0177}{\emph{PoS} {\bfseries CORFU2017} (2018) 177} [\href{https://arxiv.org/abs/1803.07670}{{\ttfamily 1803.07670}}].

\bibitem{Anastasiou:2018rdx}
A.~Anastasiou, L.~Borsten, M.J.~Duff, S.~Nagy and M.~Zoccali, \emph{{Gravity as Gauge Theory Squared: A Ghost Story}}, \href{https://doi.org/10.1103/PhysRevLett.121.211601}{\emph{Phys. Rev. Lett.} {\bfseries 121} (2018) 211601} [\href{https://arxiv.org/abs/1807.02486}{{\ttfamily 1807.02486}}].

\bibitem{Godazgar:2022gfw}
M.~Godazgar, C.N.~Pope, A.~Saha and H.~Zhang, \emph{{BRST symmetry and the convolutional double copy}}, \href{https://doi.org/10.1007/JHEP11(2022)038}{\emph{JHEP} {\bfseries 11} (2022) 038} [\href{https://arxiv.org/abs/2208.06903}{{\ttfamily 2208.06903}}].

\bibitem{Luna:2022dxo}
A.~Luna, N.~Moynihan and C.D.~White, \emph{{Why is the Weyl double copy local in position space?}}, \href{https://doi.org/10.1007/JHEP12(2022)046}{\emph{JHEP} {\bfseries 12} (2022) 046} [\href{https://arxiv.org/abs/2208.08548}{{\ttfamily 2208.08548}}].

\bibitem{Monteiro:2021ztt}
R.~Monteiro, S.~Nagy, D.~O'Connell, D.~Peinador~Veiga and M.~Sergola, \emph{{NS-NS spacetimes from amplitudes}}, \href{https://doi.org/10.1007/JHEP06(2022)021}{\emph{JHEP} {\bfseries 06} (2022) 021} [\href{https://arxiv.org/abs/2112.08336}{{\ttfamily 2112.08336}}].

\bibitem{Keeler:2024bdt}
C.~Keeler and N.~Monga, \emph{{On type-II Spacetimes and the Double Copy for Fluids Metrics}},  \href{https://arxiv.org/abs/2404.03195}{{\ttfamily 2404.03195}}.

\bibitem{penrose1976any}
R.~Penrose, \emph{Any space-time has a plane wave as a limit},  in \emph{Differential geometry and relativity}, pp.~271--275, Springer (1976).

\bibitem{Blau:2006ar}
M.~Blau, D.~Frank and S.~Weiss, \emph{{Fermi coordinates and Penrose limits}}, \href{https://doi.org/10.1088/0264-9381/23/11/020}{\emph{Class. Quant. Grav.} {\bfseries 23} (2006) 3993} [\href{https://arxiv.org/abs/hep-th/0603109}{{\ttfamily hep-th/0603109}}].

\bibitem{Chawla:2024mse}
S.~Chawla, K.~Fransen and C.~Keeler, \emph{{The Penrose limit of the Weyl double copy}}, \href{https://doi.org/10.1088/1361-6382/ad8f8c}{\emph{Class. Quant. Grav.} {\bfseries 41} (2024) 245015} [\href{https://arxiv.org/abs/2406.14601}{{\ttfamily 2406.14601}}].

\bibitem{Penrose:1985bww}
R.~Penrose and W.~Rindler, \emph{{Spinors and Space-Time}}, Cambridge Monographs on Mathematical Physics, Cambridge Univ. Press, Cambridge, UK (4, 2011), \href{https://doi.org/10.1017/CBO9780511564048}{10.1017/CBO9780511564048}.

\bibitem{Fackerell:1972hg}
E.D.~Fackerell and J.R.~Ipser, \emph{{Weak electromagnetic fields around a rotating black hole}}, \href{https://doi.org/10.1103/PhysRevD.5.2455}{\emph{Phys. Rev. D} {\bfseries 5} (1972) 2455}.

\bibitem{li1979expansions}
W.-Q.~Li and W.-T.~Ni, \emph{Expansions of the affinity, metric and geodesic equations in fermi normal coordinates about a geodesic}, {\emph{Journal of Mathematical Physics} {\bfseries 20} (1979) 1925}.

\bibitem{geroch1973space}
R.~Geroch, A.~Held and R.~Penrose, \emph{A space-time calculus based on pairs of null directions}, {\emph{Journal of Mathematical Physics} {\bfseries 14} (1973) 874}.

\bibitem{Kunze:2004qd}
K.E.~Kunze, \emph{{Behavior of curvature and matter in the Penrose limit}}, \href{https://doi.org/10.1103/PhysRevD.71.063518}{\emph{Phys. Rev. D} {\bfseries 71} (2005) 063518} [\href{https://arxiv.org/abs/gr-qc/0411115}{{\ttfamily gr-qc/0411115}}].

\bibitem{Gueven:2000ru}
R.~Gueven, \emph{{Plane wave limits and T duality}}, \href{https://doi.org/10.1016/S0370-2693(00)00517-7}{\emph{Phys. Lett. B} {\bfseries 482} (2000) 255} [\href{https://arxiv.org/abs/hep-th/0005061}{{\ttfamily hep-th/0005061}}].

\bibitem{Tod:2019urw}
P.~Tod, \emph{{Spacetimes with all Penrose limits diagonalisable}}, \href{https://doi.org/10.1088/1361-6382/ab738a}{\emph{Class. Quant. Grav.} {\bfseries 37} (2020) 075021} [\href{https://arxiv.org/abs/1909.07756}{{\ttfamily 1909.07756}}].

\bibitem{Adamo:2021dfg}
T.~Adamo and U.~Kol, \emph{{Classical double copy at null infinity}}, \href{https://doi.org/10.1088/1361-6382/ac635e}{\emph{Class. Quant. Grav.} {\bfseries 39} (2022) 105007} [\href{https://arxiv.org/abs/2109.07832}{{\ttfamily 2109.07832}}].

\bibitem{Cardoso:2016amd}
G.~Cardoso, S.~Nagy and S.~Nampuri, \emph{{Multi-centered $ \mathcal{N}=2 $ BPS black holes: a double copy description}}, \href{https://doi.org/10.1007/JHEP04(2017)037}{\emph{JHEP} {\bfseries 04} (2017) 037} [\href{https://arxiv.org/abs/1611.04409}{{\ttfamily 1611.04409}}].

\bibitem{Boulware:1968zz}
D.G.~Boulware and L.S.~Brown, \emph{{Tree Graphs and Classical Fields}}, \href{https://doi.org/10.1103/PhysRev.172.1628}{\emph{Phys. Rev.} {\bfseries 172} (1968) 1628}.

\bibitem{Duff:1973zz}
M.J.~Duff, \emph{{Quantum Tree Graphs and the Schwarzschild Solution}}, \href{https://doi.org/10.1103/PhysRevD.7.2317}{\emph{Phys. Rev. D} {\bfseries 7} (1973) 2317}.

\bibitem{Neill:2013wsa}
D.~Neill and I.Z.~Rothstein, \emph{{Classical Space-Times from the S Matrix}}, \href{https://doi.org/10.1016/j.nuclphysb.2013.09.007}{\emph{Nucl. Phys. B} {\bfseries 877} (2013) 177} [\href{https://arxiv.org/abs/1304.7263}{{\ttfamily 1304.7263}}].

\bibitem{Mougiakakos:2020laz}
S.~Mougiakakos and P.~Vanhove, \emph{{Schwarzschild-Tangherlini metric from scattering amplitudes in various dimensions}}, \href{https://doi.org/10.1103/PhysRevD.103.026001}{\emph{Phys. Rev. D} {\bfseries 103} (2021) 026001} [\href{https://arxiv.org/abs/2010.08882}{{\ttfamily 2010.08882}}].

\bibitem{DiVecchia:2023frv}
P.~Di~Vecchia, C.~Heissenberg, R.~Russo and G.~Veneziano, \emph{{The gravitational eikonal: From particle, string and brane collisions to black-hole encounters}}, \href{https://doi.org/10.1016/j.physrep.2024.06.002}{\emph{Phys. Rept.} {\bfseries 1083} (2024) 1} [\href{https://arxiv.org/abs/2306.16488}{{\ttfamily 2306.16488}}].

\bibitem{Adamo:2021hno}
T.~Adamo, A.~Ilderton and A.J.~MacLeod, \emph{{One-loop multicollinear limits from 2-point amplitudes on self-dual backgrounds}}, \href{https://doi.org/10.1007/JHEP12(2021)207}{\emph{JHEP} {\bfseries 12} (2021) 207} [\href{https://arxiv.org/abs/2103.12850}{{\ttfamily 2103.12850}}].

\bibitem{Berenstein:2002jq}
D.E.~Berenstein, J.M.~Maldacena and H.S.~Nastase, \emph{{Strings in flat space and pp waves from N=4 superYang-Mills}}, \href{https://doi.org/10.1088/1126-6708/2002/04/013}{\emph{JHEP} {\bfseries 04} (2002) 013} [\href{https://arxiv.org/abs/hep-th/0202021}{{\ttfamily hep-th/0202021}}].

\bibitem{Callan:2003xr}
C.G.~Callan, Jr., H.K.~Lee, T.~McLoughlin, J.H.~Schwarz, I.~Swanson and X.~Wu, \emph{{Quantizing string theory in AdS(5) x S**5: Beyond the pp wave}}, \href{https://doi.org/10.1016/j.nuclphysb.2003.09.008}{\emph{Nucl. Phys. B} {\bfseries 673} (2003) 3} [\href{https://arxiv.org/abs/hep-th/0307032}{{\ttfamily hep-th/0307032}}].

\bibitem{Callan:2004uv}
C.G.~Callan, Jr., T.~McLoughlin and I.~Swanson, \emph{{Holography beyond the Penrose limit}}, \href{https://doi.org/10.1016/j.nuclphysb.2004.06.033}{\emph{Nucl. Phys. B} {\bfseries 694} (2004) 115} [\href{https://arxiv.org/abs/hep-th/0404007}{{\ttfamily hep-th/0404007}}].

\bibitem{McLoughlin:2004dh}
T.~McLoughlin and I.~Swanson, \emph{{N-impurity superstring spectra near the pp-wave limit}}, \href{https://doi.org/10.1016/j.nuclphysb.2004.09.025}{\emph{Nucl. Phys. B} {\bfseries 702} (2004) 86} [\href{https://arxiv.org/abs/hep-th/0407240}{{\ttfamily hep-th/0407240}}].

\bibitem{Minahan:2005mx}
J.A.~Minahan, A.~Tirziu and A.A.~Tseytlin, \emph{{1/J corrections to semiclassical AdS/CFT states from quantum Landau-Lifshitz model}}, \href{https://doi.org/10.1016/j.nuclphysb.2005.12.003}{\emph{Nucl. Phys. B} {\bfseries 735} (2006) 127} [\href{https://arxiv.org/abs/hep-th/0509071}{{\ttfamily hep-th/0509071}}].

\bibitem{Minahan:2005qj}
J.A.~Minahan, A.~Tirziu and A.A.~Tseytlin, \emph{{1/J**2 corrections to BMN energies from the quantum long range Landau-Lifshitz model}}, \href{https://doi.org/10.1088/1126-6708/2005/11/031}{\emph{JHEP} {\bfseries 11} (2005) 031} [\href{https://arxiv.org/abs/hep-th/0510080}{{\ttfamily hep-th/0510080}}].

\bibitem{Kawai:1985xq}
H.~Kawai, D.C.~Lewellen and S.H.H.~Tye, \emph{{A Relation Between Tree Amplitudes of Closed and Open Strings}}, \href{https://doi.org/10.1016/0550-3213(86)90362-7}{\emph{Nucl. Phys. B} {\bfseries 269} (1986) 1}.

\bibitem{Beisert:2010jr}
N.~Beisert et~al., \emph{{Review of AdS/CFT Integrability: An Overview}}, \href{https://doi.org/10.1007/s11005-011-0529-2}{\emph{Lett. Math. Phys.} {\bfseries 99} (2012) 3} [\href{https://arxiv.org/abs/1012.3982}{{\ttfamily 1012.3982}}].

\bibitem{Asano:2015qwa}
Y.~Asano, D.~Kawai, H.~Kyono and K.~Yoshida, \emph{{Chaotic strings in a near Penrose limit of AdS$_{5} \times$ T$^{1,1}$}}, \href{https://doi.org/10.1007/JHEP08(2015)060}{\emph{JHEP} {\bfseries 08} (2015) 060} [\href{https://arxiv.org/abs/1505.07583}{{\ttfamily 1505.07583}}].

\bibitem{Alencar:2021ljc}
G.~Alencar and M.O.~Tahim, \emph{{Non-integrability of strings in $AdS_{6}\times S^{2}\times \Sigma $ background and its 5D holographic duals}}, \href{https://doi.org/10.1140/epjc/s10052-023-11225-3}{\emph{Eur. Phys. J. C} {\bfseries 83} (2023) 189} [\href{https://arxiv.org/abs/2106.11288}{{\ttfamily 2106.11288}}].

\bibitem{McLoughlin:2022jyt}
T.~McLoughlin and A.~Spiering, \emph{{Chaotic spin chains in AdS/CFT}}, \href{https://doi.org/10.1007/JHEP09(2022)240}{\emph{JHEP} {\bfseries 09} (2022) 240} [\href{https://arxiv.org/abs/2202.12075}{{\ttfamily 2202.12075}}].

\bibitem{Fransen:2023eqj}
K.~Fransen, \emph{{Quasinormal modes from Penrose limits}}, \href{https://doi.org/10.1088/1361-6382/acf26d}{\emph{Class. Quant. Grav.} {\bfseries 40} (2023) 205004} [\href{https://arxiv.org/abs/2301.06999}{{\ttfamily 2301.06999}}].

\bibitem{Kapec:2024lnr}
D.~Kapec and A.~Sheta, \emph{{pp-Waves and the Hidden Symmetries of Black Hole Quasinormal Modes}},  \href{https://arxiv.org/abs/2412.08551}{{\ttfamily 2412.08551}}.

\bibitem{Bhagwat:2021kwv}
S.~Bhagwat, C.~Pacilio, E.~Barausse and P.~Pani, \emph{{Landscape of massive black-hole spectroscopy with LISA and the Einstein Telescope}}, \href{https://doi.org/10.1103/PhysRevD.105.124063}{\emph{Phys. Rev. D} {\bfseries 105} (2022) 124063} [\href{https://arxiv.org/abs/2201.00023}{{\ttfamily 2201.00023}}].

\bibitem{Bhagwat:2023jwv}
S.~Bhagwat, C.~Pacilio, P.~Pani and M.~Mapelli, \emph{{Landscape of stellar-mass black-hole spectroscopy with third-generation gravitational-wave detectors}}, \href{https://doi.org/10.1103/PhysRevD.108.043019}{\emph{Phys. Rev. D} {\bfseries 108} (2023) 043019} [\href{https://arxiv.org/abs/2304.02283}{{\ttfamily 2304.02283}}].

\bibitem{Pitte:2024zbi}
C.~Pitte, Q.~Baghi, M.~Besan\c{c}on and A.~Petiteau, \emph{{Exploring tests of the no-hair theorem with LISA}}, \href{https://doi.org/10.1103/PhysRevD.110.104003}{\emph{Phys. Rev. D} {\bfseries 110} (2024) 104003} [\href{https://arxiv.org/abs/2406.14552}{{\ttfamily 2406.14552}}].

\bibitem{Stephani:2003tm}
H.~Stephani, D.~Kramer, M.A.H.~MacCallum, C.~Hoenselaers and E.~Herlt, \emph{{Exact solutions of Einstein's field equations}}, Cambridge Monographs on Mathematical Physics, Cambridge Univ. Press, Cambridge (2003), \href{https://doi.org/10.1017/CBO9780511535185}{10.1017/CBO9780511535185}.

\bibitem{blau2011plane}
M.~Blau, \emph{Plane waves and penrose limits}, {\emph{Lecture Notes for the ICTP School on Mathematics in String and Field Theory (June 2-13 2003)} (2011) }.

\bibitem{Teukolsky:1973ha}
S.A.~Teukolsky, \emph{{Perturbations of a rotating black hole. 1. Fundamental equations for gravitational electromagnetic and neutrino field perturbations}}, \href{https://doi.org/10.1086/152444}{\emph{Astrophys. J.} {\bfseries 185} (1973) 635}.

\end{thebibliography}\endgroup

\end{document}